\documentclass{article}

\usepackage{arxiv}

\usepackage[utf8]{inputenc} 
\usepackage[T1]{fontenc}    
\usepackage{hyperref}       
\usepackage{url}            
\usepackage{booktabs}       
\usepackage{amsfonts}       
\usepackage{nicefrac}       
\usepackage{microtype}      
\usepackage{graphicx}
\usepackage{doi}

\usepackage{xspace}
\usepackage[compress,numbers]{natbib}
\usepackage{float}
\usepackage{subcaption}
\usepackage{caption}
\usepackage{tabularray}
\usepackage{stfloats}

\usepackage{amsmath}

\def\tsc#1{\csdef{#1}{\textsc{\lowercase{#1}}\xspace}}
\tsc{WGM}
\tsc{QE}
\newcommand{\etal}{\textit{et al}.\xspace}

\newcommand{\ie}{i.\,e.\xspace}

\title{Electrode SOC and SOH estimation with electrode-level ECMs}

\date{}

\author{ \href{https://orcid.org/0000-0003-4967-9619}{\includegraphics[scale=0.06]{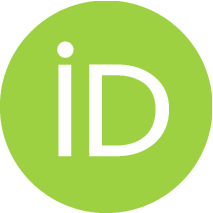}\hspace{1mm}Iker Lopetegi}  \\
	Electronic and Computer Science Department\\
	Mondragon Unibertsitatea\\
	Arrasate, Basque Country \\
	\texttt{ilopetegui@mondragon.edu} \\
	\And
	\href{https://orcid.org/0009-0006-2385-6743}{\includegraphics[scale=0.06]{orcid.eps}\hspace{1mm}Sergio Fernandez} \\
	Electronic and Computer Science Department\\
	Mondragon Unibertsitatea\\
	Arrasate, Basque Country \\
	\texttt{sergio.fernandez@alumni.mondragon.edu} \\
	\AND
     \href{https://orcid.org/0000-0002-4868-7863}{\includegraphics[scale=0.06]{orcid.eps}\hspace{1mm}Gregory L. Plett} \\
	Department of Electrical and Computer Engineering\\
    University of Colorado Colorado Springs \\
	Colorado Springs, CO 80918, United States \\
	\texttt{gplett@uccs.edu} \\
	\And
	\href{https://orcid.org/0000-0001-7991-024X}{\includegraphics[scale=0.06]{orcid.eps}\hspace{1mm}M. Scott Trimboli} \\
	Department of Electrical and Computer Engineering\\
    University of Colorado Colorado Springs \\
	Colorado Springs, CO 80918, United States \\
	\texttt{mtrimbol@uccs.edu} \\
	\And
	\href{https://orcid.org/0000-0003-4012-9426}{\includegraphics[scale=0.06]{orcid.eps}\hspace{1mm}Unai Iraola} \\
	Electronic and Computer Science Department\\
	Mondragon Unibertsitatea\\
	Arrasate, Basque Country \\
	\texttt{uiraola@mondragon.edu} \\
}

\renewcommand{\headeright}{}
\renewcommand{\undertitle}{}
\renewcommand{\shorttitle}{Electrode SOC and SOH estimation with electrode-level ECMs}

\begin{document}
\maketitle

\begin{abstract}
	Being able to predict battery internal states that are related to battery degradation is a key aspect to improve battery lifetime and performance, enhancing cleaner electric transportation and energy generation. However, most present battery management systems (BMSs) use equivalent-circuit models (ECMs) for state of charge (SOC) and state of health (SOH) estimation. These models are not able to predict these aging-related variables, and therefore, they cannot be used to limit battery degradation. In this paper, we propose a method for electrode-level SOC (eSOC) and electrode-level SOH (eSOH) estimation using an electrode-level ECM (eECM). The method can produce estimates of the states of lithiation (SOL) of both electrodes and update the eSOH parameters to maintain estimation accuracy through the lifetime of the battery. Furthermore, the eSOH parameter estimates are used to obtain degradation mode information, which could be used to improve state estimation, health diagnosis and prognosis. The method was validated in simulation and experimentally.
\end{abstract}


\section{Introduction}

The rapid shift towards renewable energy sources, and the need to electrify the transport sector, has heightened the need for advanced energy storage solutions. Lithium-ion (Li-ion) batteries are superior to other chemistries in terms of energy density, lifespan, self-discharge and efficiency \cite{khan2024state}. In this context, accurate Li-ion battery modelling plays a key role in optimizing their performance, extending their life cycle and ensuring security \cite{WANG2023100260}. A battery model helps in predicting its behavior under different power profiles and external conditions, improving operation \cite{Plett2016,Plett2015b}.
Battery models are composed of mathematical equations to emulate (usually) the battery voltage response to a current input under given external conditions, such as temperature or pressure. These mathematical equations may contain internal battery states that cannot be measured with conventional sensors. Two of these pivotal internal states are the state of charge (SOC) and state of health (SOH). The SOC represents the remaining capacity until full discharge, and the SOH represents the capacity of the battery with respect to its nominal value.

Different Li-ion battery modelling approaches have been proposed in the literature, such as empirical, electrochemical or data-driven models. However, equivalent-circuit models (ECMs) have been adopted most widely by industry. ECMs are behavioral models that represent the battery dynamics by using an electrical circuit analogy. The complexity of these models can vary, as presented in \cite{nikdel2014various}, but the adopted standard in the literature and in industry for different Li-ion technologies is an electrical circuit composed by a controlled voltage source, an equivalent series resistance (ESR) and one or more parallel resistor-capacitor (RC) branches in series \cite{tran2021comparative}. The controlled voltage source is known as open circuit voltage (OCV), which models the battery voltage when no current is flowing and the cell is in equilibrium. The ESR models the instantaneous voltage drop or increase when the battery is connected to a load or a charger, respectively. The RC branches model diffusion voltages which can be approximated closely adding more RC sub circuits in series. The elements presented above are modeled mainly as a function of SOC (temperature and current sign dependence may also be considered). Thus, model performance relies on accurate SOC estimation. Overall, these models have shown a good complexity-accuracy trade-off in battery voltage response, SOC and SOH estimation, as was shown by Plett \cite{plett2004extended,plett2004extended2,plett2004extended3, plett2006sigma,plett2006sigma2}. Nevertheless, in spite of having reduced computational and conceptual complexity, these models do not offer any information that can predict battery aging. Therefore, they could not be used to prevent degradation mechanisms that affect the performance and safety of lithium-ion batteries \cite{Plett2015b}.

Consequently, models that give more information about battery internal dynamics are being investigated in the literature. Physics-based models (PBMs) could be an alternative to ECMs due to their ability to give internal physical information \cite{Plett2015b}. These models could be used for advanced state estimation and control \cite{wang2023system,plett2024battery,lopetegi2024,lopetegi2024b}, however, they have two main drawbacks: the computational and conceptual complexity is higher than in ECMs, and the parametrization process is more time-consuming and expensive \cite{plett2024battery}. Many efforts have been made in the last years to reduce the computational cost of these models \cite{rodriguez2019comparing}, and some of them have been implemented in embedded systems such as BMSs \cite{Sturm2019,Miguel2021}. The parametrization of these models is also being investigated in the literature \cite{oca2021physico,Rojas2024,lu2021pulse,lu2022eis}. However, due to the mentioned inconveniences, these models have not yet been adopted by the industry.

With the intention of bridging the gap between these two approaches, some authors have proposed a mid-term solution for state estimation and control: electrode-level ECMs (eECMs) \cite{zhang2022real,zhao2021study,drees2021fast}. These models describe battery electrode behavior and their relation using an electric circuit analogy for both the positive electrode (PE) and negative electrode (NE). They have similar computational cost to ECMs, but are able to predict separate positive and negative electrode potentials, which could be used for fast charging algorithm design; for example, as was done by Drees \etal \cite{drees2021fast}. Zhao \etal \cite{zhao2021study} used pseudo-2-dimensional (P2D) model simulations to develop an eECM. Zhang \etal \cite{zhang2022real} developed an eECM using a reference electrode to obtain electrode potential measurements and estimated both electrodes' state of lithiation (SOL) and potentials using an extended Kalman filter (EKF). These works are promising for future BMS implementation, since they offer more information than traditional ECMs with similar computational and conceptual cost. However, none of the above-mentioned studies investigated on SOH estimation using these models. 

Several detrimental side-reactions can occur in lithium-ion batteries \cite{han2019review,Edge2021}. These reactions change the behavior of the battery, and could make the state estimation algorithm behave poorly unless model parameter values are adjusted \cite{Plett2016,lopetegi2024,lopetegi2024b}. Thus, it is necessary to take into account battery degradation in order to maintain accuracy in the SOC estimate \cite{Plett2016}. Lithium-ion battery aging can be classified into three main degradation modes: loss of active material (LAM) in both electrodes and the loss of lithium inventory (LLI) \cite{birkl2017degradation}. These three degradation modes change the shape of the OCV, generating an error on the SOC estimation in traditional ECM based algorithms and even in most PBM-based algorithms proposed in the literature \cite{lopetegi2024,lopetegi2024b}. Thus, efforts have been made in estimating electrode-level SOH (eSOH) parameters and, therefore, all the degradation modes \cite{Lee2018,Mohtat2019,lee2020electrode,SFernandez2024ModEst}. The importance of eSOH estimation for accurate SOC estimation was highlighted in \cite{lopetegi2024}. However, most of the methods presented in the literature require low C-rate data and are not ideal for real-world applications. In our previous work \cite{lopetegi2024,lopetegi2024b}, we developed a state estimator capable of estimating all the eSOH parameters using a PBM and interconnected Sigma-Point Kalman Filters (SPKFs), which could work without low C-rate data. Nevertheless, since it is based on a PBM, it is probably not ideal from the industry point of view for its implementation at the moment.

Compared to the previous works that we have cited, the method proposed in this paper makes significant contributions: (i) it obtains all the eSOH parameters together with eSOC estimates using an eECM and without requiring low current tests or additional sensors. To the best of the authors' knowledge, this is the first method that is able to estimate all the eSOH parameters and the electrode-level SOC with an eECM; (ii) it proposes a parametrization method that takes into account bigger electrode lithiation ranges than the usually considered SOC range; (iii) it obtains degradation mode values, which are important for battery diagnosis and prognosis \cite{costa2024icformer}, using an eECM. 

The outline of the paper is as follows: First, we describe the modeling approach and the parametrization. Next, the state and parameter estimation algorithm is discussed. Afterwards, the proposed method is validated and the results obtained with the proposed method are shown and discussed. Finally, we present the conclusions drawn from this work.


\section{Model development}

The eECM development is covered in this section. The modeled battery is an LG M50 with NMC/graphite-SiO\textsubscript{x} chemistry. The model equations are presented first. Later, the characterization process to obtain the model parameter values is discussed.

\subsection{Electrode-level ECM}

A constant temperature second-order ECM output voltage equation is defined as:
\begin{equation}\label{eq:ECMoutput}
    v(t) = OCV(z(t)) - v_{C_{1}}(t) - v_{C_{2}}(t) - R_{0}i(t),
\end{equation}
where $z$ is the SOC, $R_0$ is the ESR, and $v_{C_{1}}$ and $v_{C_{2}}$ represent the voltages across the RC networks, which are calculated as follows:
\begin{equation}\label{eq:VCcalc}
    \dot{v}_{C_{n}}(t) = \frac{1}{C_n}\left(i(t) - \frac{v_{C_{n}}(t)}{R_n}\right).
\end{equation}

\begin{figure}[htb!]
    \centering
    \subfloat[\centering\label{fig:Model:ECMcircuit}]{{\includegraphics[width=0.4\linewidth]{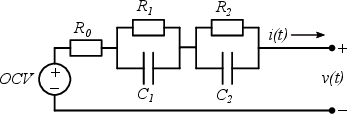}}}\\
    \subfloat[\centering\label{fig:Model:HCcircuit}]{{\includegraphics[width=0.4\linewidth]{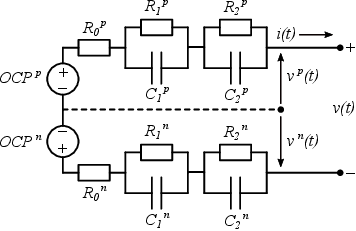}}}
    \caption{Electrical circuit representation. (a) Traditional ECM; (b) HC model.}
    \label{fig:Model:ECMvsHC}
\end{figure}

An eECM model is conceptually similar to an ECM, as illustrated in Figure \ref{fig:Model:ECMvsHC}. The output potential of the PE is given by
\begin{equation}\label{eq:Vpos}
     v^p(t) = OCP^{p}(\theta^{p}(t)) - v^p_{C_{1}}(t) - v^p_{C_{2}}(t) - R^p_{0}i(t),
\end{equation}
where $v^p_{C_{1}}(t)$ and $v^p_{C_{2}}(t)$ are the voltages of the capacitors, $R^p_{0}$ is the series resistance of the positive electrode, and $\theta^p$ is the SOL of the positive electrode. The NE potential is given by
\begin{equation}\label{eq:Vneg}
     v^n(t) = OCP^{n}(\theta^{n}(t)) + v^n_{C_{1}}(t) + v^n_{C_{2}}(t) + R^n_{0}i(t),
\end{equation}
where $v^n_{C_{1}}(t)$ and $v^n_{C_{2}}(t)$ are the voltages of the capacitors, $R^n_{0}$ is the series resistance of the negative electrode, and $\theta^n$ is the SOL of the negative electrode. The total cell voltage is calculated as:
\begin{equation}\label{eq:Vcell}
      v(t) = v^p(t) - v^n(t).
\end{equation}

Summing up, each electrode is represented by its own half-cell (HC) ECM, allowing the calculation of individual electrode potentials. The OCP values depend on the SOL of each electrode, which can be modeled with the coulomb counting method. Additionally, the passive circuit elements capture the behavior of each electrode, enabling the decoupling of voltage polarization and diffusion dynamics for both electrodes.

\subsection{Model parametrization}

The passive circuit elements of an eECM can be characterized as a function of SOC \cite{zhang2022real}, similar to what is typically done in ECMs. This approach can give accurate results when the cell is in its beginning-of-life (BOL) condition, before any aging has occurred. However, as cells age, the stoichiometry ranges of the electrodes between 100\% and 0\% SOC change due to LLI and LAM \cite{birkl2017degradation}. These degradation modes will generate a shift between the electrodes, and their capacities will decrease, meaning that a single SOC condition may correspond to different SOL values at BOL and any other aged condition. Consequently, the performed characterization, which assumes fixed stoichiometric windows, may no longer accurately represent the actual behavior of the battery. Thus, in this work, we propose an alternative characterization method for the passive circuit elements of each electrode as a function of the electrode SOL. This allows for a decoupled characterization that should reflect more accurately the electrode behavior when batteries age.

\begin{figure}[htb!]
    \centering
    \subfloat[\centering\label{fig:ParamFit:HPPCpos_I}]{{\includegraphics[width=0.4\linewidth]{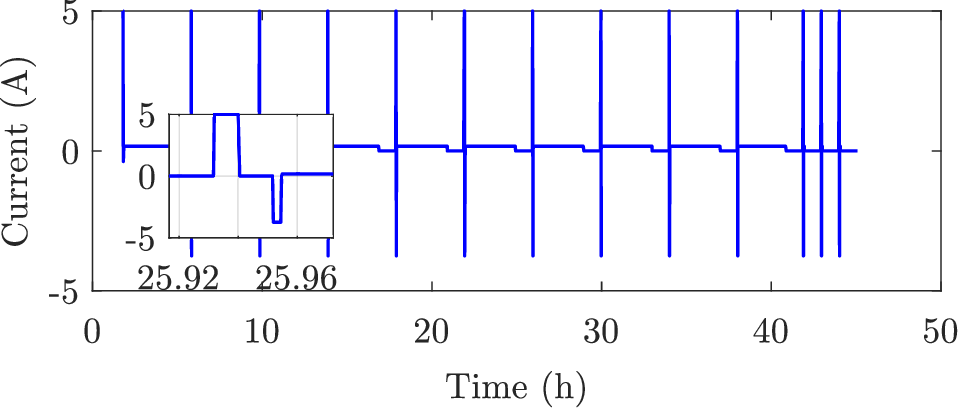}}}\\
    \subfloat[\centering\label{fig:Model:HPPCpos_SOL}]{{\includegraphics[width=0.4\linewidth]{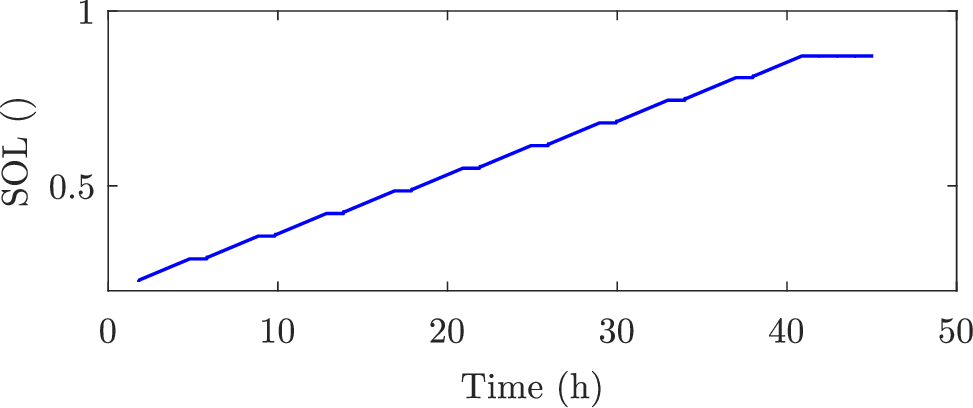}}}
    \caption{HPPC test. (a) Cell current profile; (b) PE SOL evolution during the test.}
    \label{fig:ParamFit:HPPCpos}
\end{figure}

To achieve this, a Hybrid Power Pulse Characterization (HPPC) test is performed at the electrode level, in a sufficiently wide SOL range to cover aged cases, as shown in Figure \ref{fig:ParamFit:HPPCpos}. A parameter estimation algorithm is then used to fit the values of the passive circuit elements as a function of SOL. This method presents two main experimental challenges. First, if a reference electrode is inserted into the cell to obtain the individual electrode potentials, as in \cite{zhang2022real}, wide stoichiometry ranges cannot be characterized, since extreme cell voltage values required for such a characterization will deteriorate the cell, leading to inaccurate results. An alternative approach could be to build half cells from the electrode materials, as for OCP obtention tests, and to perform HPPC tests on them. Nevertheless, the impedances of these reconstructed cells will differ significantly from those of the original full cell, as has been reported in the literature \cite{son2024}.

Therefore, in this paper, we have adopted an alternative approach. We performed HPPC test simulations using the Single Particle Model with electrolyte dynamics (SPMe) presented in \cite{lopetegi2023}, to obtain both electrode potential data separately. This method allows us to decouple the impedance contributions of each electrode and characterize the eECM impedance over a wider stoichiometry range. The PBM parameters were acquired from \cite{chen2020}, and are presented in Table \ref{tab:PBMparams}. Besides, the OCP functions that were used in both the PBM and in the eECMs are defined in Appendix \ref{ap:PBMparams}. Since cell to cell variations may make some parameters inaccurate for our experimental LG M50 cells, certain parameters of the SPMe were optimized. The diffusion coefficients were updated in a previous work \cite{yeregui2023state} to improve the voltage response compared to our experimental cells. Furthermore, the stoichiometric windows were adapted to improve the voltage response. All the parameters can be found in Table \ref{tab:PBMparams}.

\begin{figure}[htb!]
    \centering
    \subfloat[\centering\label{fig:ParamFit:Vpos}]{{\includegraphics[width=0.4\linewidth]{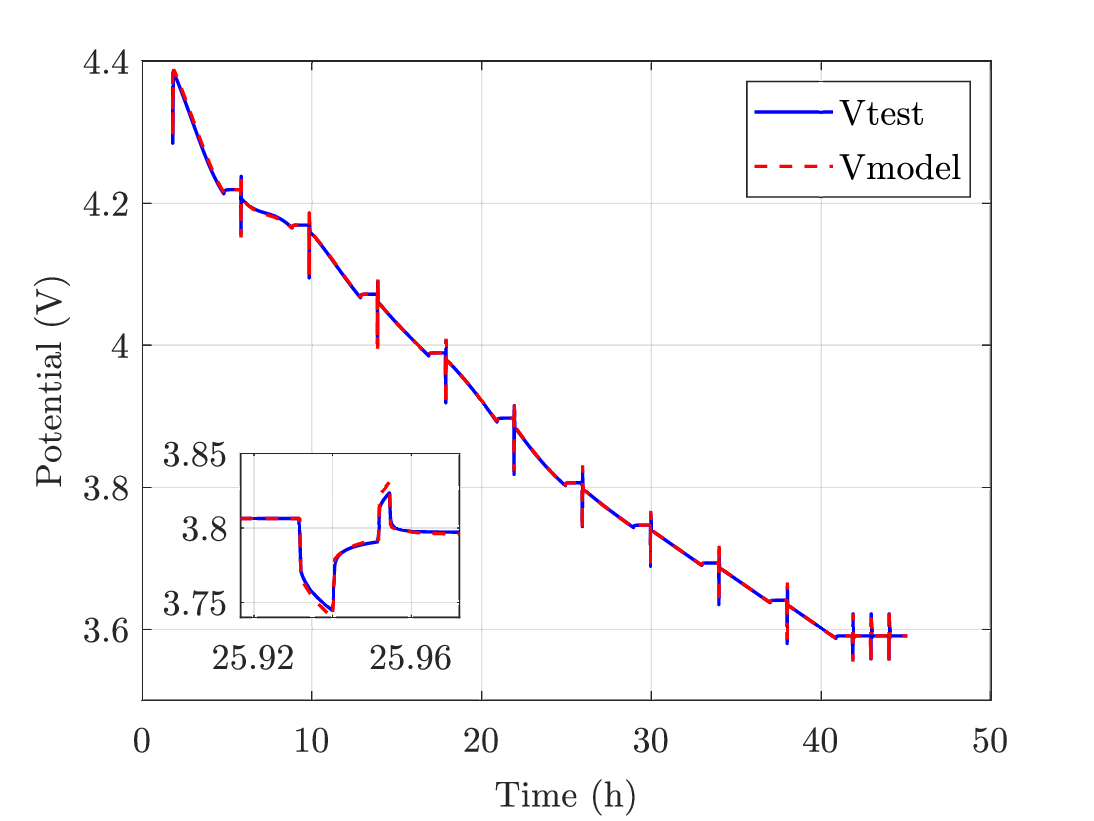}}}\\
    \subfloat[\centering\label{fig:ParamFit:DVpos}]{{\includegraphics[width=0.4\linewidth]{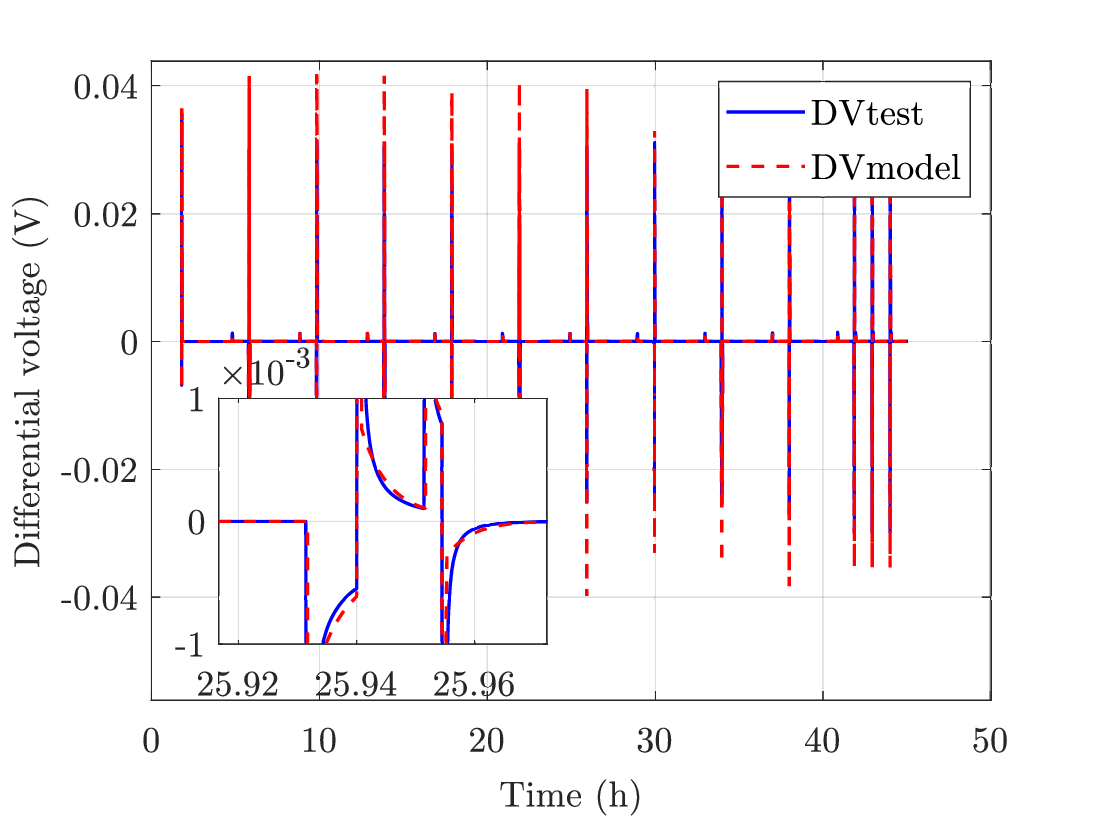}}}
    \caption{PE parameter estimation results. (a) Model voltage and experimental voltage comparison; (b) Differential voltage response comparison.}
    \label{fig:ParamFitOutput_pos}
\end{figure}

The parameter estimation algorithm for the passive circuit elements of the HC models employs Matlab's built-in genetic algorithm (GA) for the optimization. The optimization is initially performed as a single-objective process, using the following cost function:
\begin{equation} \label{eq:J1}
    J_1 = \sqrt{\frac{1}{n}\sum^{n}_{i=1}(V_i-V_i^*)^2},
\end{equation}
where $V_i$ is the measured test voltage, and $V_i^*$ is the model output voltage. Although this approach yields reasonably good results, we introduce a second cost function to improve the model fitting. By adding this second cost function, we transition from single-objective to multi-objective optimization. The additional cost function is defined as
\begin{equation} \label{eq:J2}
    J_2 = \sqrt{\frac{1}{n}\sum^{n}_{i=1}\left(\frac{\Delta V_i - \Delta V_i^*}{\Delta t_i}\right)^2},
\end{equation}
where $\Delta V_i$ is the measured test differential voltage and $\Delta V_i^*$ is the model output differential voltage. Differential voltages are defined as the difference between consecutive voltage samples ($\Delta V_i = V_i - V_{i-1}$). This second cost function allows the optimizer to more accurately fit the voltage response during rest periods, ensuring the model captures both fast and slow dynamic behavior of the system. In Figure \ref{fig:ParamFitOutput_pos}, the results obtained from the optimization of the passive elements for the PE are shown. The results for the NE passive elements are presented in Appendix \ref{ap:NE_parametriztion}.

The Root Mean Square (RMS) errors obtained for the PE voltage response and differential voltage response are 1.30 mV and 0.18 mV, respectively. For the NE, the RMS error for the voltage response is 5.10 mV, and 2.50 mV for the differential voltage response. The optimized values for the passive circuit elements are presented in Table \ref{tab:HCparams}.

\begin{table*}[htb!]
\centering
\caption{\label{tab:HCparams} eECM passive circuit element values.}
\label{tab:params}
\resizebox{\textwidth}{!}{%
    \begin{tabular}{@{}*{11}{l}}
    \toprule  
    SOL ($\theta$) (\%) & $R_0^n (\text{m}\Omega)$ & $R_1^n (\text{m}\Omega)$ & $C_1^n (\text{k}F)$ & $R_2^n (\text{m}\Omega)$ & $C_2^n (\text{k}F)$ & $R_0^p (\text{m}\Omega)$ & $R_1^p (\text{m}\Omega)$ & $C_1^p (\text{k}F)$ & $R_2^p (\text{m}\Omega)$ & $C_2^p (\text{k}F)$ \cr 
        \midrule
    0 & 128 & 0.05 & 3259 & 0.03 & 5596& 2.7& 3.1& 20.3847& 1.8& 75.2961\cr
    10 & 25.7 & 0.08 & 1915 & 0.08 & 2329& 11.8 & 9.8& 3.7071& 12.2& 9.311\cr
    20 & 26.1 & 5.1 & 6.85 & 0.47& 200.2& 4.1 & 14.3& 1.1344& 14.2& 0.7966\cr
    30 & 23.7 & 5.5 & 21.2 & 3& 25.65 & 7.4 & 14.2& 13.9334& 1.9& 1.6941\cr
    40 & 22 & 5.3 & 12.61 & 6 & 12.81 & 8.1& 14.2& 12.2308& 4.3& 3.8551\cr
    50 & 22.4 & 7.3 & 6.7 & 1.4 & 75.74& 7.6& 12.4& 14.4738& 7.5& 5.7076\cr
    60 & 22.6 & 9.4 & 9.22 & 2.1 & 51.98 & 8& 14.2& 13.9521& 3.5& 5.279\cr
    70 & 24.6 & 3.5 & 15.8& 5.9 & 9.89 & 5.9& 11.5& 15.1133& 1.7& 4.3032\cr
    80 & 23.8 & 2.9 & 7.92 & 4.7 & 15.52 & 7.5& 8.7& 20.4305& 3& 3.5433\cr
    90 & 34.2 & 5.8 & 15.4 & 4.2& 18.92 & 6.6& 10.3& 6.7177& 4.5& 16.2523\cr
    100 & 32.2 & 5.2& 0.84 & 8.3& 11.04& 9& 5& 32.348& 8.1& 23.688\cr           
    \bottomrule
    \end{tabular}}
\end{table*}

\subsection{Model validation}

\begin{figure}[htb!]
    \centering
    \includegraphics[width=0.4\linewidth]{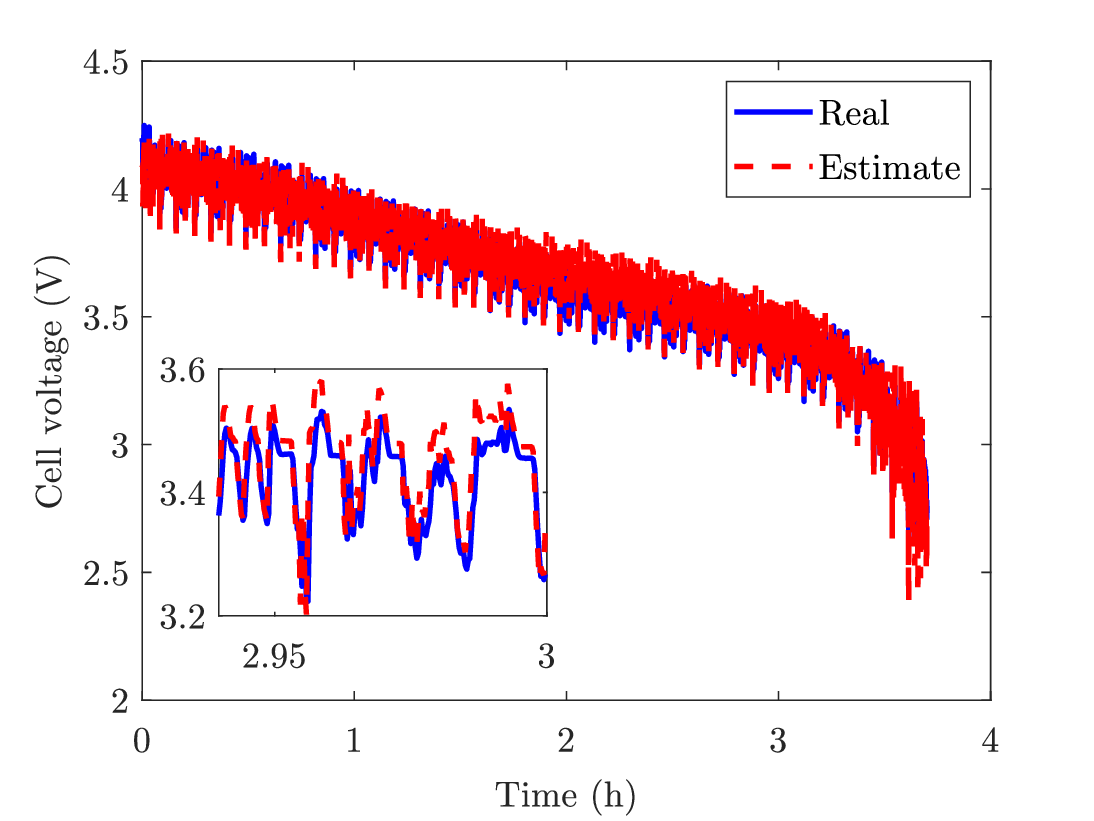}
    \caption{Experimental cell voltage and model voltage response comparison for a discharge with consecutive UDDS cycles.}
    \label{fig:Model:UDDSoutput}
\end{figure}

Both the OCPs and the impedances of the electrodes are then combined to construct the full-cell model. The model is validated using a dynamic current profile, different from the one used for characterization. In this case, an Urban Dynamometer Driving Schedule (UDDS) was employed for validation. The model output compared to the experimental cell voltage is presented in Figure \ref{fig:Model:UDDSoutput}. The RMS error of the voltage prediction is 14 mV.

\section{State and parameter estimator}

Due to the nonlinear nature of lithium-ion batteries, nonlinear state estimation algorithms are usually employed to estimate the internal states of the battery \cite{plett2004extended,plett2006sigma,Plett2016}. The most widely used state estimation algorithm is probably the EKF \cite{plett2004extended,Bizeray2015,Stetzel2015,zhang2022real}, which relies on linearization to propagate the system's mean and covariance. The EKF has been used extensively to estimate the SOLs and potentials of the electrodes \cite{Bizeray2015,Stetzel2015,couto2016soc,zhang2022real}. However, the linearization can lead to unreliable estimates when the system exhibits significant nonlinearity \cite{Simon2006}.

In contrast, the sigma point Kalman filter (SPKF) can (should) provide better covariance estimates by propagating the mean and covariance through a set of function evaluations \cite{Simon2006}, thereby avoiding the need for linearization. Thus, we preferred to use the SPKF, as it generally produces more accurate results than the EKF \cite{Simon2006}.

Many state estimation algorithms such as EKF or SPKF, or state observer algorithms, have been shown to give accurate results in the literature for SOC and SOH estimation \cite{Plett2016}; however, when estimating both electrodes' concentrations, a weak observability issue arises \cite{Allam2018a}. Both electrodes' contribution to cell voltage must be estimated from just one cell voltage measurement, making the estimation of both variables challenging. To improve the observability of these variables, Allam and Onori \cite{Allam2018a} proposed to use an interconnected observer approach. The idea behind the method relies on the claim that the electrode concentrations are observable from cell voltage measurements if the other state variables are known, which was demonstrated by Bartlett \etal \cite{Bartlett2016}. In the interconnected observer architecture, one of the observers estimates the states of the positive electrode while maintaining the negative electrode in open-loop, and the other
observer estimates the states of the negative electrode with the positive electrode in open-loop. The estimates are interconnected, feeding the open-loop model of the other observer, correcting the error of each open-loop model \cite{Allam2018a}. Since this approach was presented, it has been used in many other works to improve the observability of the state variables \cite{lopetegi2024,lopetegi2024b,Allam2018a,Allam2020a,dey2019battery,sattarzadeh2020addressing,deSouza2024interconnected}.

\subsection{The interconnected SPKF}

The eECM exhibits the same weak observability issue with respect to estimating both electrodes' SOLs. Therefore, an interconnected SPKF architecture was adopted, similar to that employed in \cite{lopetegi2024}.

For the state estimation algorithm, we first reformulate our model in a general nonlinear form as
\begin{gather}
    \dot{x} = f(x,u) \nonumber \\
     y = h(x,u),
\end{gather}
where $x$ represents the state variables of the system, $u$ represents the inputs, $y$ is the output, $f(\cdot)$ is a nonlinear state transition function that describes how the state evolves and $h(\cdot)$ is a nonlinear function that maps the system state to the output.

As can be seen in the eECM output Eq. \ref{eq:Vcell}, the electrode potential equations \ref{eq:Vpos} and \ref{eq:Vneg} depend on the state variables $v_{C_{1}}$, $v_{C_{2}}$ and $\theta$. Thus, the state $x$ is divided into two parts, $x_p$ and $x_n$:
\begin{gather}
    x_p = \left( \begin{matrix} v^p_{C_{1}} \\
    v^p_{C_{2}}\\
    \theta^{p} \end{matrix} \right), \nonumber \\
    x_n = \left( \begin{matrix} v^n_{C_{1}} \\
    v^n_{C_{2}}\\
    \theta^{n} \end{matrix} \right).
\end{gather}

The state estimator is thus composed of two SPKFs, which are interconnected with each other. The estimator itself can be divided into two main phases: the prediction phase and the correction phase. The prediction phase starts by taking the two state and covariance estimates of the previous time step. Then, the two state-space systems are used to obtain the state and covariance predictions. Afterwards, all the predictions, which are based on the two separated estimates of the previous time step, are interconnected to obtain two output predictions. In the correction phase, the prediction information is
compared with the cell voltage measurement and used to obtain the filter gains. Then, these gains are used to compute the two state and covariance estimates.

A detailed description of the six recursive steps of the estimator is given below:

\textit{Step 1a: State prediction time update}. In this step, the new state is predicted based on previous information. The standard SPKF algorithm uses sigma points of the state, process noise, and measurement noise vectors to represent the randomness of the system. However, since the state equations of our model are linear, we can simplify the state prediction to,
\begin{gather}
    \hat{x}_p^-[k] = A_p'\hat{x}_p^+[k-1] + B_p'u[k-1] \nonumber \\
    \hat{x}_n^-[k] = A_n'\hat{x}_n^+[k-1] + B_n'u[k-1],
\end{gather}
where $\hat{x}^+$ is the state estimate, $\hat{x}^-$ the state prediction, and $k$ and $k-1$ denote the present and previous time steps, respectively. $A_p'$, $B_p'$, $A_n'$ and $B_n'$ are the discrete-time form of $A_p$, $B_p$, $A_n$ and $B_n$ state-space matrices which are defined, based on Eq. \ref{eq:Vpos} and Eq. \ref{eq:Vneg}, as follows:
\begin{gather} \label{eq:state-space_matrices}
    A_r' = \left( \begin{matrix} -\frac{1}{R^r_1 C^r_1} & 0 & 0\\ 
    0 & -\frac{1}{R^r_2 C^r_2} & 0\\
    0 & 0 & 1 \end{matrix} \right) \nonumber \\
    B_r' = \left( \begin{matrix} \frac{1}{C^r_1}\\
    \frac{1}{C^r_2}\\
    \frac{\eta^r}{3600 Q^r} \end{matrix} \right) \nonumber,
\end{gather}
where $\eta$ is the coulombic efficiency and $r$ denotes negative or positive electrode components.

\textit{Step 1b: Error covariance time update}. The error covariance is defined by
\begin{equation}
    \Sigma_x^-[k] = \mathbb{E}[(x[k] - \hat{x}^-[k])((x[k] - \hat{x}^-[k])^T],
\end{equation}
which leads to the two covariance matrices
\begin{gather}
    \Sigma_{\tilde{x}_{p}}^-[k] = A_p'\Sigma_{\tilde{x}_{p}}^+[k-1](A_p')^T + \Sigma_{\tilde{w}_{p}}, \nonumber \\
    \Sigma_{\tilde{x}_{n}}^-[k] = A_n'\Sigma_{\tilde{x}_{n}}^+[k-1](A_n')^T + \Sigma_{\tilde{w}_{n}}.
\end{gather}

\textit{Step 1c: Output prediction}. Since the output equation is nonlinear, the sigma points $\mathcal{X}_p^-[k]$ and $\mathcal{X}_n^-[k]$ must be calculated first. The sets of sigma points for each estimator are calculated as
\begin{gather}
    \mathcal{X}_{p} ^- [k] = \left\{ \hat{x}_p^- [k], \hat{x}_p^- [k] + h \sqrt{\Sigma_{\tilde{x}_p}^-}, \hat{x}_p^- [k] - h \sqrt{\Sigma_{\tilde{x}_p}^-} \right\} \nonumber \\
     \mathcal{X}_{n} ^- [k] = \left\{ \hat{x}_n^- [k], \hat{x}_n^- [k] + h \sqrt{\Sigma_{\tilde{x}_n}^-}, \hat{x}_n^- [k] - h \sqrt{\Sigma_{\tilde{x}_n}^-} \right\},
    \label{}
\end{gather}
where $h$ is a tuning variable for the SPKF and $\sqrt{\Sigma_{\tilde{x}_p}^-}$ and $\sqrt{\Sigma_{\tilde{x}_n}^-}$ are  the lower-triangular matrix square roots of the error covariance matrices, which were computed using a Cholesky decomposition. The sigma points (the vectors of the set $\mathcal{X}$) are then used to compute the output equation and to obtain the output sigma points,
\begin{gather}
    \mathcal{Y}_{p,i} = h(\hat{x}_n^-[k],\mathcal{X}_{p,i}^- [k],u[k])\nonumber \\
    \mathcal{Y}_{n,i} = h(\mathcal{X}_{n,i}^- [k],\hat{x}_p^-[k],u[k]).
\end{gather}

The two cell voltage predictions are then calculated as the weighted
mean of these sigma points:
\begin{gather}
    \hat{y}_p[k] = \sum_{i=0}^{2N_{x_p}+1} \alpha_{p,i}^{(m)} \mathcal{Y}_{p,i}[k] \nonumber \\
    \hat{y}_n[k] = \sum_{i=0}^{2N_{x_n}+1} \alpha_{n,i}^{(m)} \mathcal{Y}_{n,i}[k]
\end{gather}
where $\alpha_{p,i}^{(m)}$ and $\alpha_{n,i}^{(m)}$ are the constants used to calculate the weighted mean.

\textit{Step 2a: Estimator gain matrix}. In order to update the predictions with the present information, the estimator gain matrix is calculated. For that, we compute the required covariance matrices:
\begin{gather}
    \Sigma_{\tilde{y}_{p}}[k] = \sum_{i = 0}^{2N_{x_p}+1}\alpha_{p,i}^{(c)} (\mathcal{Y}_{p,i} [k]-\hat{y}_{p}[k])(\mathcal{Y}_{p,i}[k]-\hat{y}_{p}[k])^T \nonumber \\
    \Sigma_{\tilde{y}_{n}}[k] = \sum_{i = 0}^{2N_{x_n}+1}\alpha_{n,i}^{(c)} (\mathcal{Y}_{n,i} [k]-\hat{y}_{n}[k])(\mathcal{Y}_{n,i}[k]-\hat{y}_{n}[k])^T,
\end{gather}

\begin{gather}
    \Sigma_{\tilde{x}_p\tilde{y}_p}^- [k] = \sum_{i = 0}^{2N_{x_p}+1}\alpha_{p,i}^{(c)} (\mathcal{X}_{p,i}^- [k]-\hat{x}_p^- [k])(\mathcal{Y}_{p,i}[k]-\hat{y}_p[k])^T \nonumber \\
    \Sigma_{\tilde{x}_n\tilde{y}_n}^- [k] = \sum_{i = 0}^{2N_{x_n}+1}\alpha_{n,i}^{(c)} (\mathcal{X}_{n,i}^- [k]-\hat{x}_n^- [k])(\mathcal{Y}_{n,i}[k]-\hat{y}_n[k])^T,
\end{gather}
where $\alpha_{p,i}^{(c)}$ and $\alpha_{n,i}^{(c)}$ are the constants used to calculate the weighted covariance.

Once these matrices are computed, we calculate the gains of the estimators:
\begin{align}
    L_p[k] = \Sigma_{\tilde{x}_p\tilde{y}_p}^- [k] (\Sigma_{\tilde{y}_p}[k])^{-1} \nonumber \\
    L_n[k] = \Sigma_{\tilde{x}_n\tilde{y}_n}^- [k] (\Sigma_{\tilde{y}_n}[k])^{-1}.
\end{align}

\textit{Step 2b: State estimate measurement update}. The state estimates are calculated as
\begin{align}
    \hat{x}_p^+[k] = \hat{x}_p^-[k] + L_p[k](y[k] - \hat{y}_p[k]) \nonumber \\
    \hat{x}_n^+[k] = \hat{x}_n^-[k] + L_n[k](y[k] - \hat{y}_n[k]).
\end{align}

\textit{Step 2c: Error covariance measurement update}. The error covariance matrices are updated by computing
\begin{align}
    \Sigma_{\tilde{x}_p}^+[k] = \Sigma_{\tilde{x}_p}^-[k] - L_p[k] \Sigma_{\tilde{y}_p}[k] \left(L_p[k]\right)^T \nonumber \\
    \Sigma_{\tilde{x}_n}^+[k] = \Sigma_{\tilde{x}_n}^-[k] - L_n[k] \Sigma_{\tilde{y}_n}[k] \left(L_n[k]\right)^T .
\end{align}

\subsection{Electrodes' Capacity Estimation}
As was stated above, in order to maintain accurate state estimates, we need to update the eSOH parameters, \ie both electrodes' capacities and stoichiometric windows. However, these parameters have low observability to cell voltage \cite{Mohtat2019,lee2020electrode,lopetegi2024b}. Cell voltage has low sensitivity to electrode capacities, as occurs when estimating the capacity of a cell from cell voltage data \cite{Plett2016}. For that, Plett proposed different methods that relied on SOC estimates and capacity change measurements to estimate cell total capacity \cite{plett2011recursive}. Adapting these algorithms to take into account electrode capacities and SOLs, instead of cell capacity and SOC, we could estimate both electrodes total capacities.


Thus, we use the approximate weighted total least squares (AWTLS) algorithm \cite{plett2011recursive} to estimate electrode capacities. The method is based on performing a linear regression with the measured capacity change data and the estimated SOC values. In our case, since we are trying to estimate the electrode capacities, we will relate them to the electrode SOL estimates.

The PE and NE SOL are related to the electrode total capacity by the equations
\begin{equation}
    \dot{\theta^p}(t) = \frac{\eta(t)i(t)}{3600Q^p}
\end{equation}
and
\begin{equation}
    \dot{\theta^n}(t) = \frac{-\eta(t)i(t)}{3600Q^n},
\end{equation}
respectively. As can be seen, considering the current $i$ positive for discharge, NE SOL decreases and PE SOL increases when discharging. When charging, the opposite occurs. The coulombic efficiency $\eta$ is assumed to be one.

In discrete-time, the SOL of each electrode is calculated as
\begin{align}
    \theta^p_{k_2} = \theta^p_{k_1} + \frac{1}{Q^p}\sum_{k=k_1}^{k_2-1}\eta_ki_k \nonumber \\
    \theta^n_{k_2} = \theta^n_{k_1} - \frac{1}{Q^n}\sum_{k=k_1}^{k_2-1}\eta_ki_k,
\end{align}
which can be rearranged to get:
\begin{align}
    \sum_{k=k_1}^{k_2-1}\eta_ki_k \nonumber = Q^p(\theta^p_{k_2} - \theta^p_{k_1}) \\
    - \sum_{k=k_1}^{k_2-1}\eta_ki_k = Q^n(\theta^n_{k_2} - \theta^n_{k_1}).
\end{align}
These equations show a linear structure $y = Qx$, where $y = \sum_{k=k_1}^{k_2-1}\eta_ki_k$ is the measured capacity change, $(\theta_{k_2} - \theta_{k_1})$ the estimated change in SOL, and $Q$ the capacity of the electrode. From this relation, each electrode's capacity estimate can be computed using a regression technique. The merit function of the AWTLS is defined as
\begin{equation}
    \mathcal{X}_{AWTLS} = \sum_{i=1}^n \frac{(y_i - \hat{Q}x_i)^2}{(1 + \hat{Q}^2)^2}\left( \frac{\hat{Q}^2}{\sigma^2_{x_i}} + \frac{1}{\sigma^2_{y_i}}\right),
\end{equation}
where $\sigma^2_{x_i}$ and $\sigma^2_{y_i}$ are the variances of the SOL estimate difference, and measured capacity change. The detailed procedure to implement the AWTLS can be found in \cite{plett2011recursive}.

\subsection{Stoichiometric window estimation}

Stoichiometry window estimation is performed using the same method presented in \cite{lopetegi2024b}. Stoichiometry limits (electrodes SOL values at minimum and maximum SOC) define the maximum and minimum lithiation in each of the electrodes in a given health state for the 100\% and 0\% SOC conditions. These limits vary when LAM or LLI occurs \cite{birkl2017degradation,SFernandez2024ModEst}. We assume that the cell capacity is equal to the useful electrode capacity (capacity between stoichiometry limits), so
\begin{equation} \label{eq:Q_electrode}
    Q \approx Q^p(\theta^p_{0\%} - \theta^p_{100\%}) \approx Q^n(\theta^n_{100\%} - \theta^n_{0\%}).
\end{equation}

If the useful capacities of both electrodes are equivalent, then the remaining useful capacities until full charge or discharge must also be the same, obtaining the next capacity equivalencies:
\begin{align} \label{eq:Q_ch_dis}
    Q^n(\theta^n - \theta^n_{0\%}) = Q^p(\theta^p_{0\%} - \theta^p) \nonumber \\
    Q^n(\theta^n_{100\%} - \theta^n) = Q^p(\theta^p - \theta^p_{100\%}).
\end{align}
Since we also know our fixed operation voltage limits, $V_{min}$ and $V_{max}$, we know that the 100\% and 0\% positive and negative stoichiometries must fulfil the following voltage equivalencies:
\begin{align} \label{eq:V_maxmin}
    OCP^p(\theta^p_{0\%}) -  OCP^n(\theta^n_{0\%}) = V_{min} \nonumber \\
    OCP^p(\theta^p_{100\%}) -  OCP^n(\theta^n_{100\%}) = V_{max}.
\end{align}

Assuming that the OCPs are fixed functions of the stoichiometries, the only unknowns of the four-equation system given in Eqs. \ref{eq:Q_ch_dis}-\ref{eq:V_maxmin}, are the stoichiometry limits at 100\% and 0\% SOC, the cell capacities and the actual SOLs. Nevertheless, as the SOLs are estimated with the interconnected SPKFs, and the electrode capacities are obtained with the AWTLS method, the system of equations can be solved to obtain the stoichiometry windows. These stoichiometric limits are not estimated every time step, since these parameters do not change that quickly. In our case, we solved the system of equations every 10000 s. Note that this system of equations may not be solved analytically due to the OCP functions' form. For this work, we used the MATLAB \textit{vpasolve} numerical solver.

It is worth noting that the assumption of identical OCP vs. stoichiometry relations may not hold true for the entire lifetime of a battery. If we consider a multi-species, multi-reaction (MSMR) model \cite{verbrugge2017thermodynamic}, we might have different aging rates in different galleries of an electrode, which could change the shape of the OCP vs. stoichiometry curve. However, despite being an interesting topic of research to improve the proposed algorithm, in this work we will assume fixed OCP curves and will work on improving the OCP models in future work.

\subsection{Degradation Mode and SOH Estimation}

Using the eSOH parameter estimates together with the SOL estimates, the three degradation modes LLI, LAM\textsuperscript{n} and LAM\textsuperscript{p} can be calculated. The LAM of each electrode, in \%, can be defined as
\begin{eqnarray} \label{eq:LAM1}
    LAM^r = \left(1 - \frac{Q^r_a}{Q_f^r} \right) \times 100
\end{eqnarray}
where $r$ denotes the positive or negative electrode, $Q^r_a$ is the total capacity of the aged electrode $r$. $Q_f^r$ is the total capacity of a fresh electrode $r$. Using the electrode capacity estimates of the AWTLS algorithm, LAM values can be computed.

The total amount of intercalated lithium in the cell at any SOC can be calculated as
\begin{eqnarray}\label{eq:n_LI}
    n_{Li} = \frac{3600}{F} \left(z^p Q^p + z^n Q^n \right),
\end{eqnarray}
where $F$ is the Faraday constant. Any $z^n$ and $z^p$ values can be chosen from the entire SOC range (both from the same SOC). Then, the LLI can be calculated as
\begin{eqnarray}\label{eq:LLI}
    LLI = \left(1- \frac{n_{Li}^a}{n_{Li}^f} \right) \times 100,
\end{eqnarray}
where $n_{Li}^a$ is the aged lithium inventory, and $n_{Li}^f$ is the fresh lithium amount in the cell. Since the SOLs and capacities have been estimated, the calculation of the LLI is straightforward.

In addition, an SOH value can be calculated by comparing the fresh cell capacity with the aged cell capacity. The available capacity of the aged cell can be obtained by substituting the estimated eSOH parameter values in the electrode capacity Eq \ref{eq:Q_electrode}. Then, the SOH of the cell can be calculated as
\begin{eqnarray} \label{eq:SOH}
    SOH = \frac{Q_a}{Q_f},
\end{eqnarray}
where $Q_a$ is the aged cell capacity calculated with the estimated eSOH parameters, and $Q_f$ is the fresh cell capacity.


\section{Results and Discussion}

The performance of the state estimator was validated using simulation and experimental data. First, the simulation study was carried out, since in this case we can surely know whether the state estimator is converging to the correct values. After confirming that the algorithm was working well in simulation, we performed experimental tests to corroborate the results in a real scenario.

\subsection{Simulation validation}
\begin{figure}[htb!]
    \centering \includegraphics[width=0.4\linewidth]{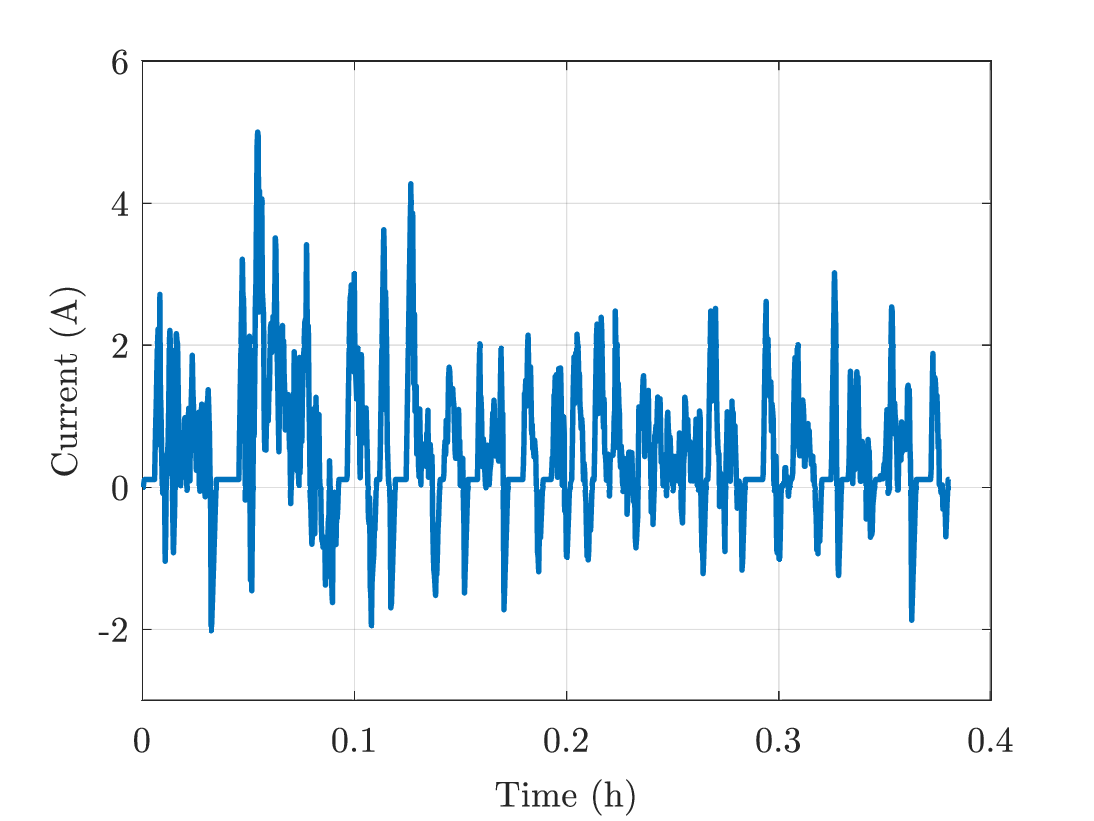}
    \caption{Charge-depleting UDDS current profile.}
    \label{fig:I_UDDS}
\end{figure}

For the simulation validation, we performed dynamic cycling simulations using a P2D model, as in \cite{lopetegi2024b}. Using a different model from the one used for characterization, a difference between the ``characterization cell'' and the ``tested cell'' is inserted. A charge-depleting UDDS profile was used to discharge the cell until the minimum voltage was reached and then it was charged at 0.5 C constant-current until reaching 4.2 V. The charge-depleting UDDS profile is shown in Figure \ref{fig:I_UDDS}.

For this test, an aged condition was simulated in the P2D model by inserting LLI and LAM in both electrodes. Initializing the eECM in the BOL condition, we can evaluate whether the estimator is able to converge or not. The inserted LAM for the positive electrode is 20\%, for the negative electrode 10\%, and the LLI is 16\%. Furthermore, the positive and negative electrode SOLs were intentionally incorrectly initialized, with a 10\% error.

\begin{figure*}[htb!]
    \centering
    \begin{subfigure}{0.32\linewidth}
        \centering
        \includegraphics[width=1\textwidth]{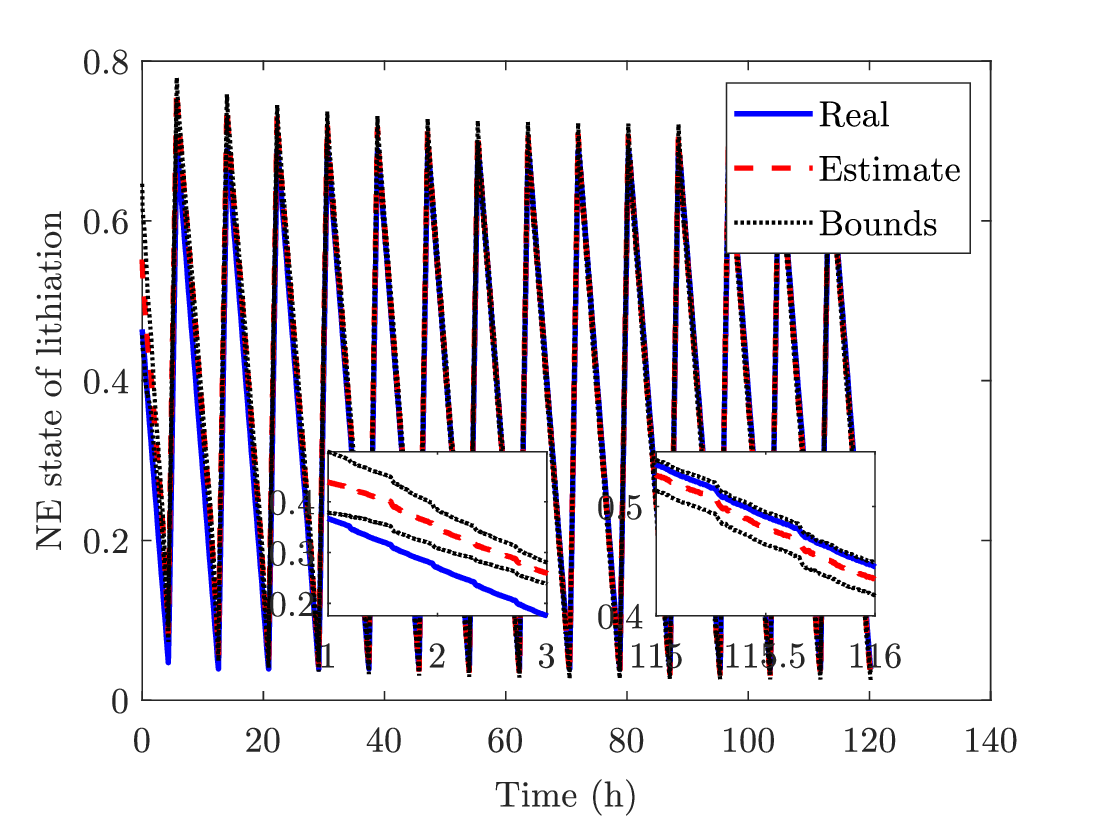}
        \caption{}\label{fig:Val_simu-SOL_n}
    \end{subfigure}
    \begin{subfigure}{0.32\linewidth}
        \centering
        \includegraphics[width=1\textwidth]{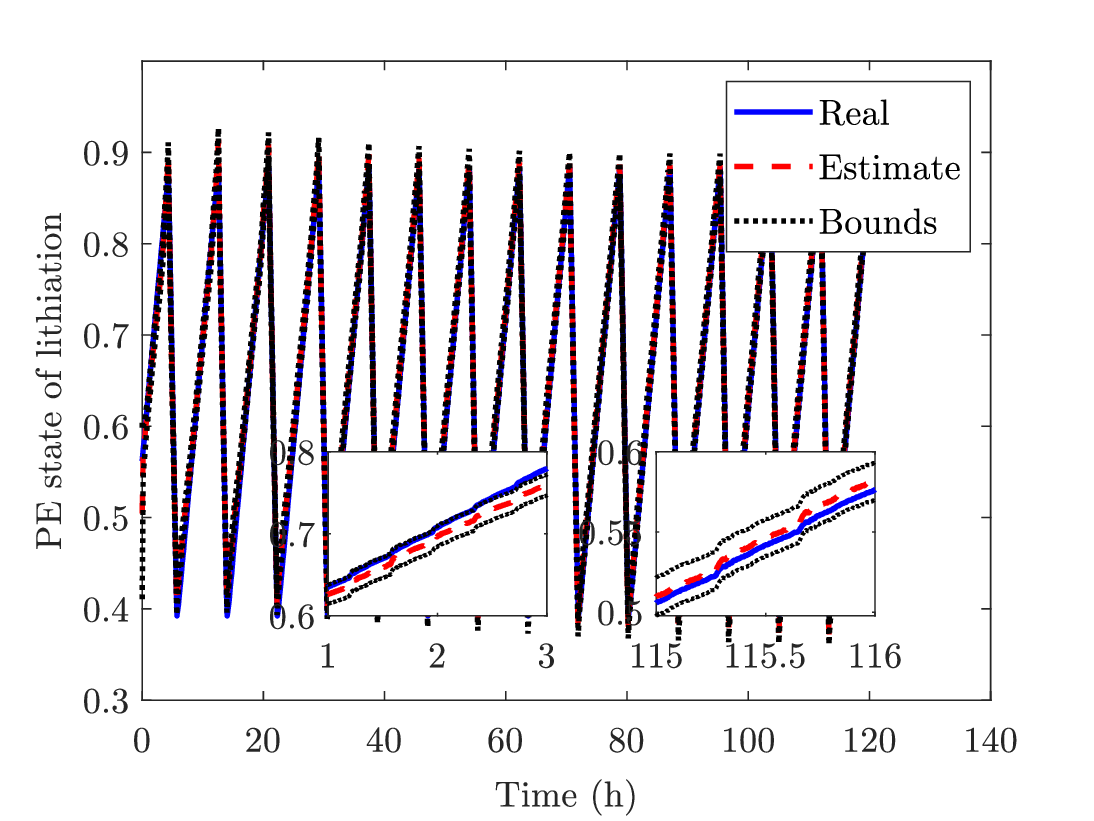}
        \caption{}\label{fig:Val_simu-SOL_p}
    \end{subfigure}
    \begin{subfigure}{0.32\linewidth}
        \centering
        \includegraphics[width=1\textwidth]{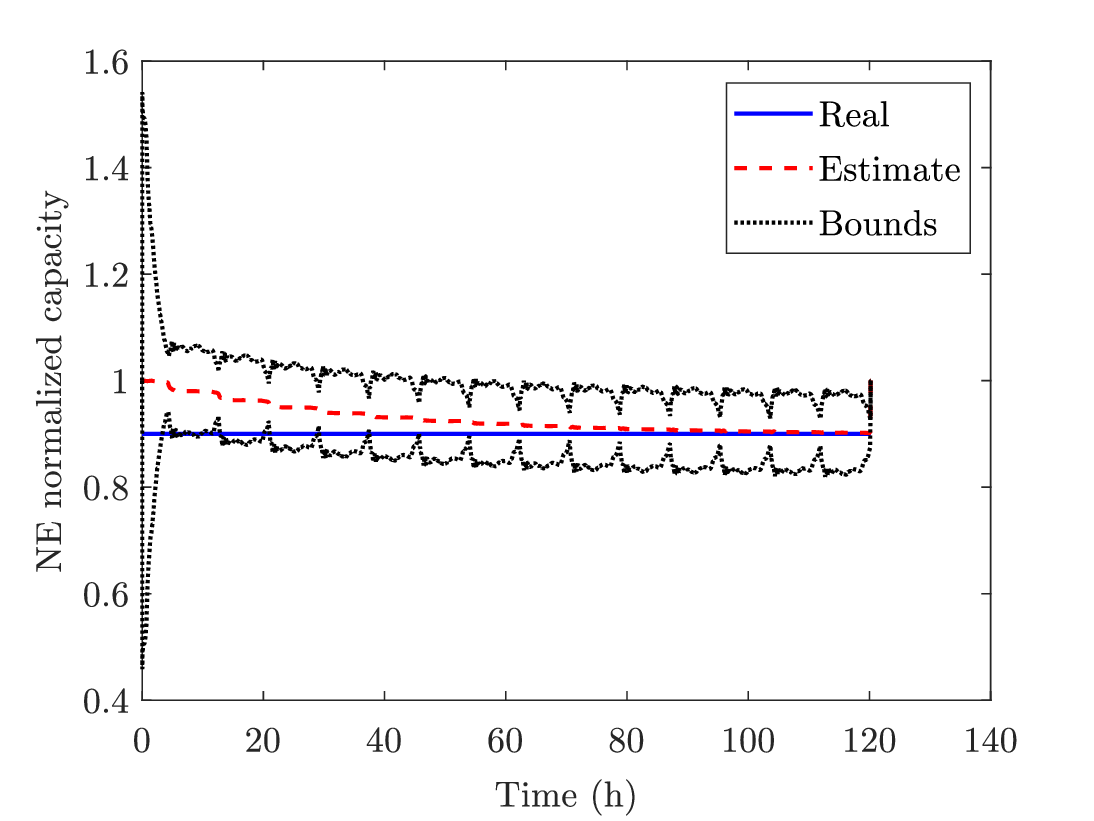}
        \caption{}\label{fig:Val_simu-SOH_n}
    \end{subfigure}
    
    \begin{subfigure}{0.32\linewidth}
        \centering
        \includegraphics[width=1\textwidth]{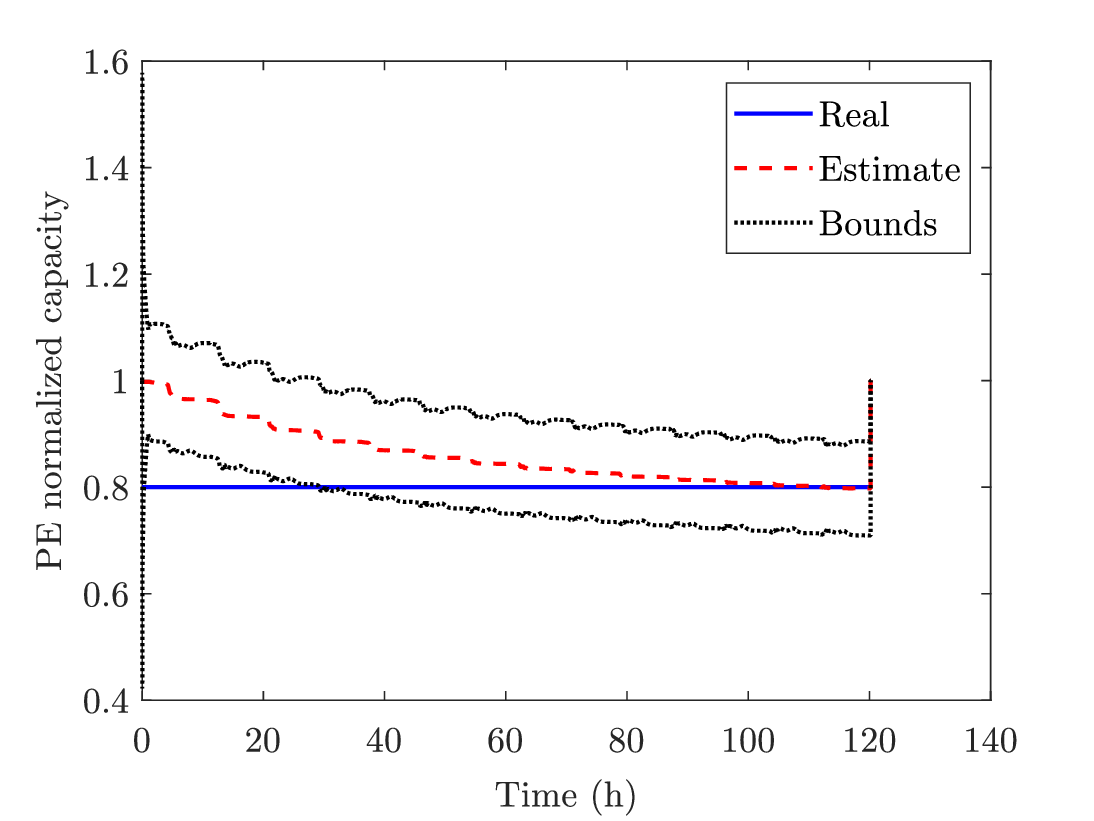}
        \caption{}\label{fig:Val_simu-SOH_p}
    \end{subfigure}
    \begin{subfigure}{0.32\linewidth}
        \centering
        \includegraphics[width=1\textwidth]{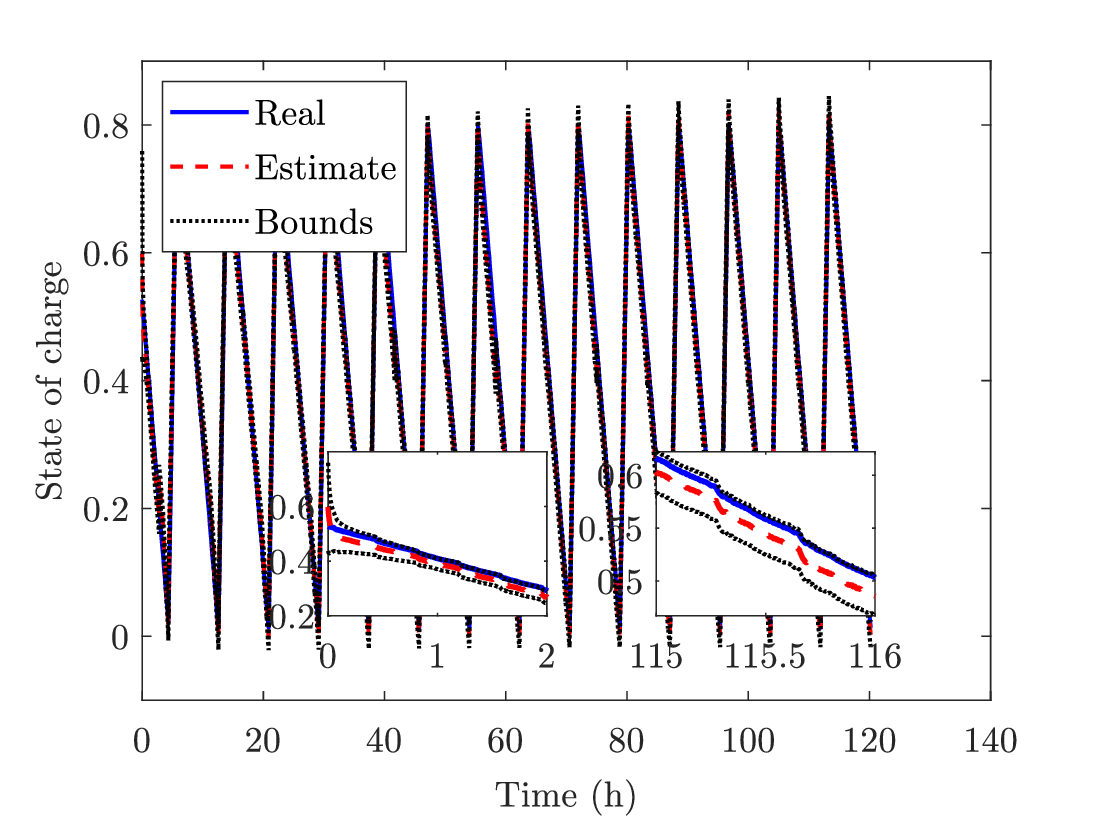}
        \caption{}\label{fig:Val_simu-SOC}
    \end{subfigure}
    \begin{subfigure}{0.32\linewidth}
        \centering
        \includegraphics[width=1\textwidth]{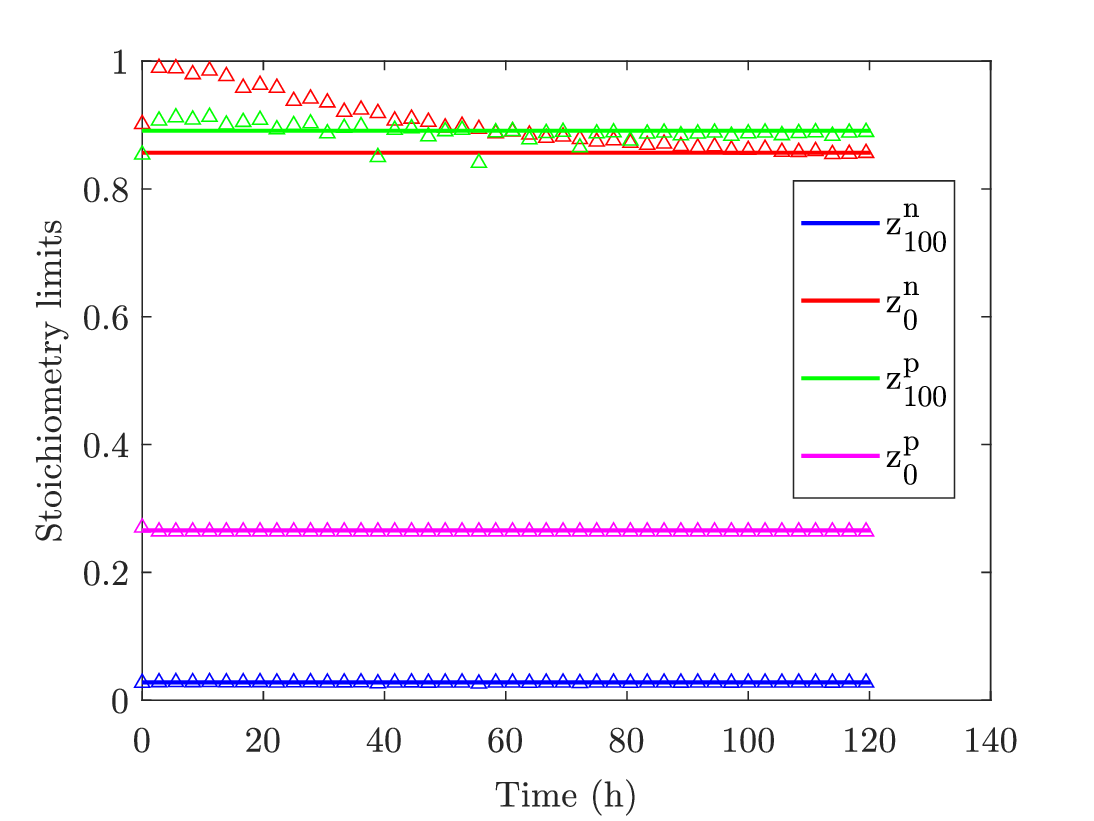}
        \caption{}\label{fig:Val_simu-Stoichiometries}
    \end{subfigure}
    
    \caption{State and parameter estimation simulation results with 10 \% LAM in the negative electrode, 20 \% LAM in the positive electrode, and 16 \% LLI. (a)-(b) Negative and positive electrode SOLs, respectively; (c) and (d) Normalized negative and positive electrode capacities, respectively; (e) SOC estimate; (f)  100 \% and 0 \% SOC stoichiometry estimates, where markers denote estimates and solid lines represent the true values.}
    \label{fig:Val_simu}
\end{figure*}

Figure \ref{fig:Val_simu} shows the results of the test. As can be observed in Figures \ref{fig:Val_simu-SOL_n} and \ref{fig:Val_simu-SOL_p}, the negative and positive electrode SOL estimates are not accurate in the first cycles, since the erroneous electrode capacity values make the state equations an inaccurate description of the SOL behavior. However, as is shown in Figures \ref{fig:Val_simu-SOH_n} and \ref{fig:Val_simu-SOH_p}, the capacities converge to the true values as the simulation progresses, correcting the SOL predictions in the state-space model, and making their estimates accurate.

The SOC estimate is shown in Figure \ref{fig:Val_simu-SOC}. As can be seen, this estimate is also inaccurate in the first cycles, since it has a direct dependence on the SOL estimate. Moreover, as the stoichiometry windows are also incorrectly estimated, as shown in Figure \ref{fig:Val_simu-Stoichiometries}, the SOL-SOC relation is not accurate. However, as the capacities and the SOLs of the electrodes are corrected, the stoichiometries are well calculated with the system of equations, making the SOC estimate also converge.

\begin{figure}[htb!]
        \centering
        \includegraphics[width=0.4\linewidth]{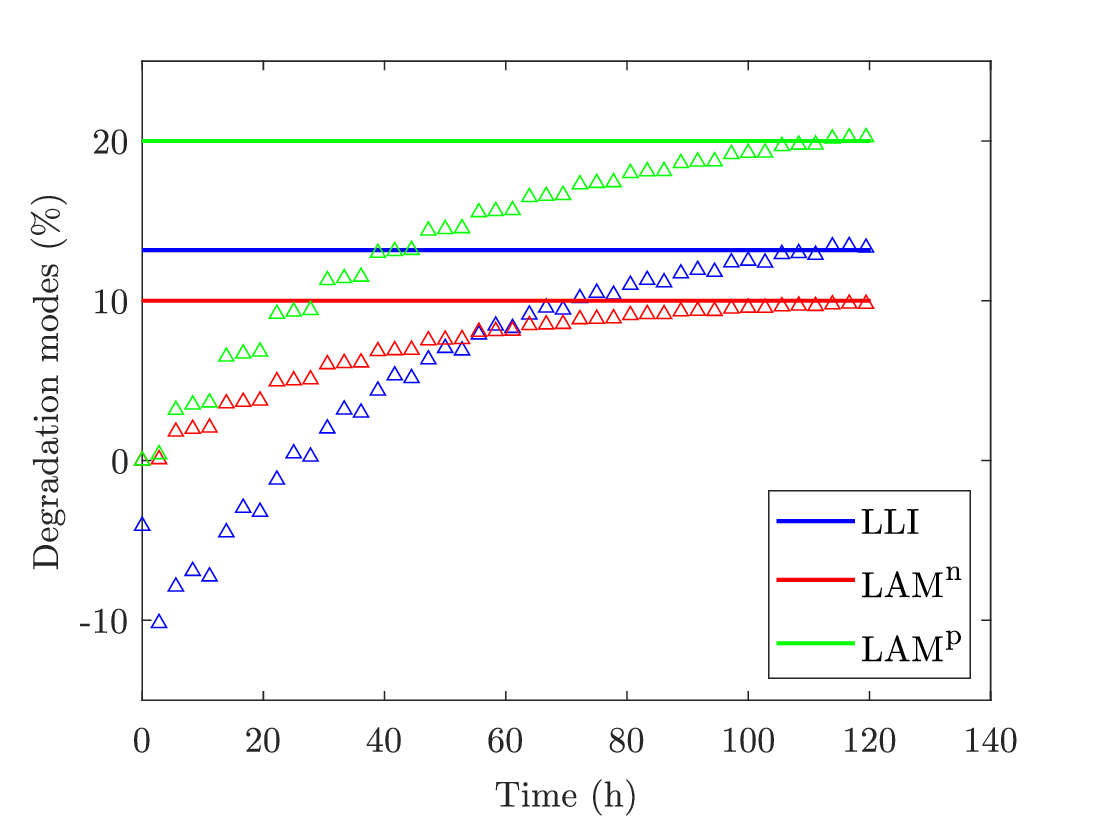}
        \caption{Degradation mode estimates from the simulation validation tests. Solid lines denote the true values and the markers the estimated values.}\label{fig:Simu_Modes}
\end{figure}

With the estimated states and parameters, \ie the SOLs and the eSOH parameters, we can compute degradation mode values for health diagnosis and prognosis purposes. The degradation mode estimates are shown in Figure \ref{fig:Simu_Modes}. As can be observed, the estimates are incorrectly calculated in the start of the simulation, but converge accurately toward the end.

\subsection{Experimental validation}
The experimental validation was divided into two parts. First, we performed driving cycle profile experiments with a fresh cell, since in this case the model parameters will be more accurate, and the estimation should be easier. By initializing the states and parameters from incorrect values, we evaluated the performance of the algorithm. Second, we took an aged cell and estimated its degradation modes from OCV data. This way, we can evaluate if the state estimation algorithm finds the correct aging modes that the cell suffered.

\subsubsection{Validation at BOL}
\begin{figure}[htb!]
        \centering
        \includegraphics[width=0.4\linewidth]{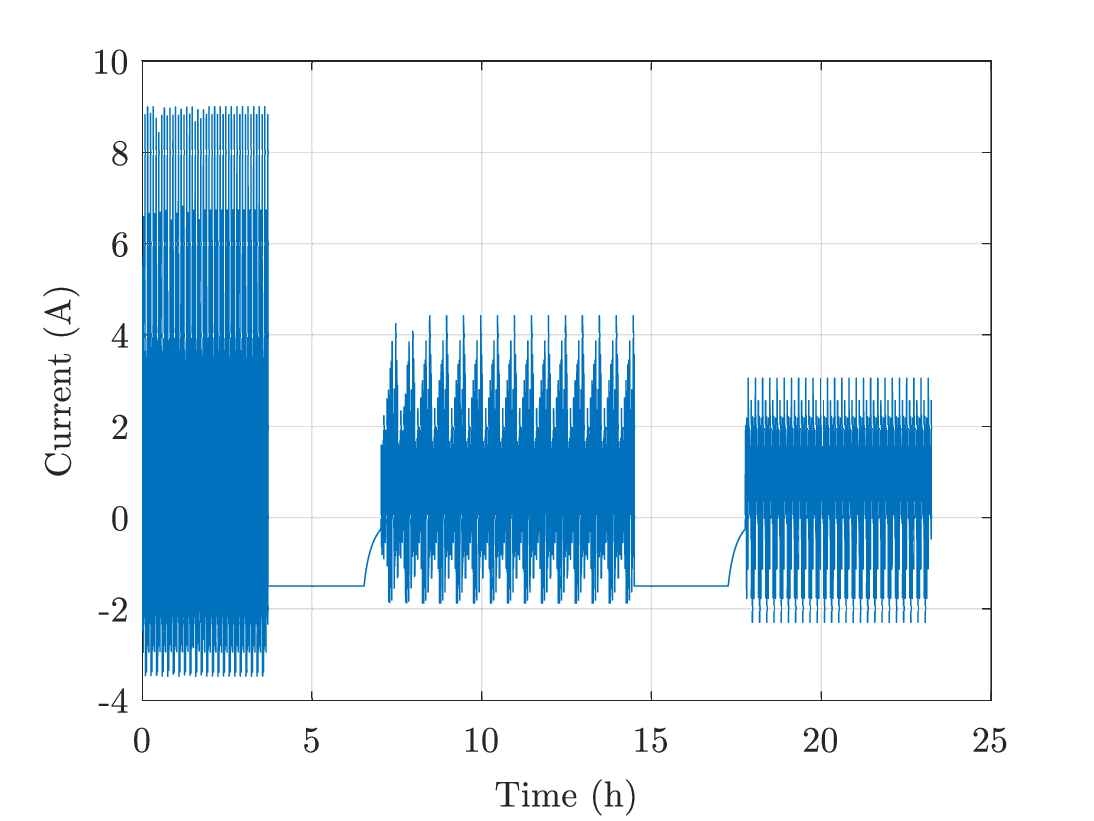}
        \caption{Experimental current profile applied in the BOL validation test.}\label{fig:EXP_I}
\end{figure}

The BOL validation was performed using previously obtained cycling data for an LG M50 cell, which was used in \cite{yeregui2023state}. The current profile used for the experiments is composed of US06, WLTC and HWFET consecutive profiles, which can be seen in Figure \ref{fig:EXP_I}. The experiments were performed at 25 ºC with a Basytec XCTS cycler.

\begin{figure*}[htb!]
    \centering
    \begin{subfigure}{0.32\linewidth}
        \centering
        \includegraphics[width=1\textwidth]{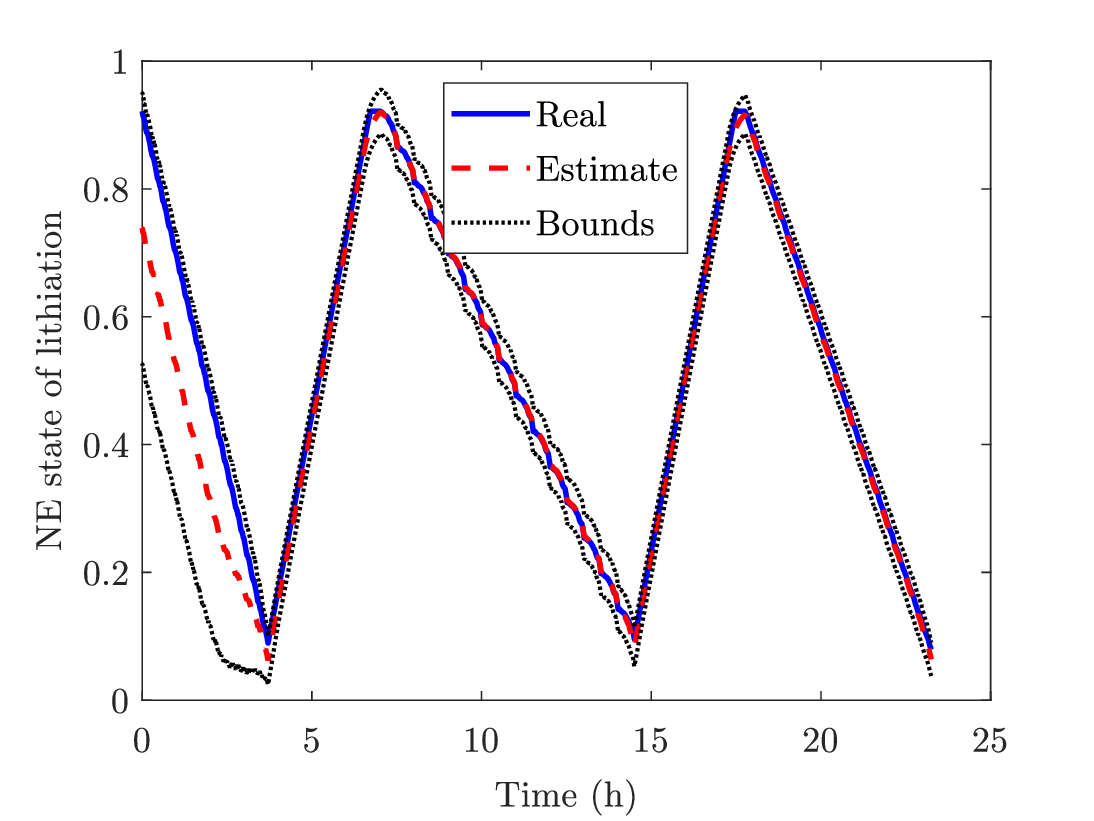}
        \caption{}\label{fig:Val_Exp-SOL_n}
    \end{subfigure}
    \begin{subfigure}{0.32\linewidth}
        \centering
        \includegraphics[width=1\textwidth]{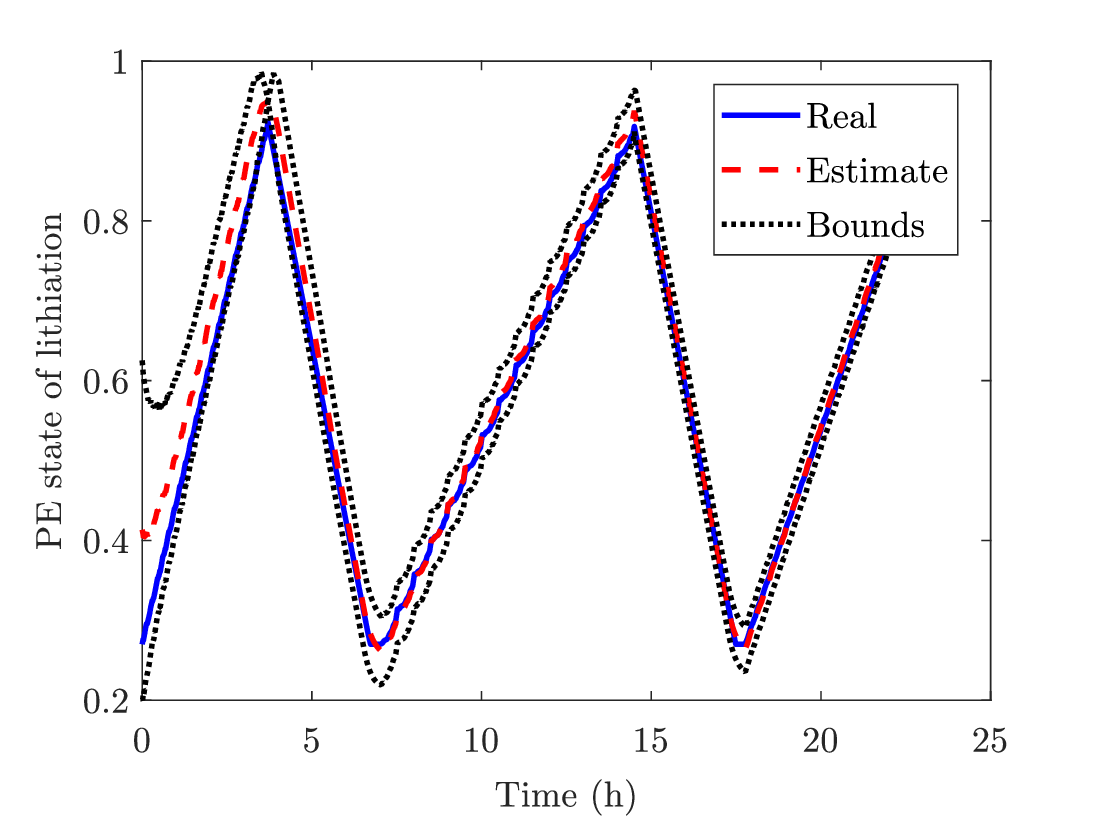}
        \caption{}\label{fig:Val_Exp-SOL_p}
    \end{subfigure}
    \begin{subfigure}{0.32\linewidth}
        \centering
        \includegraphics[width=1\textwidth]{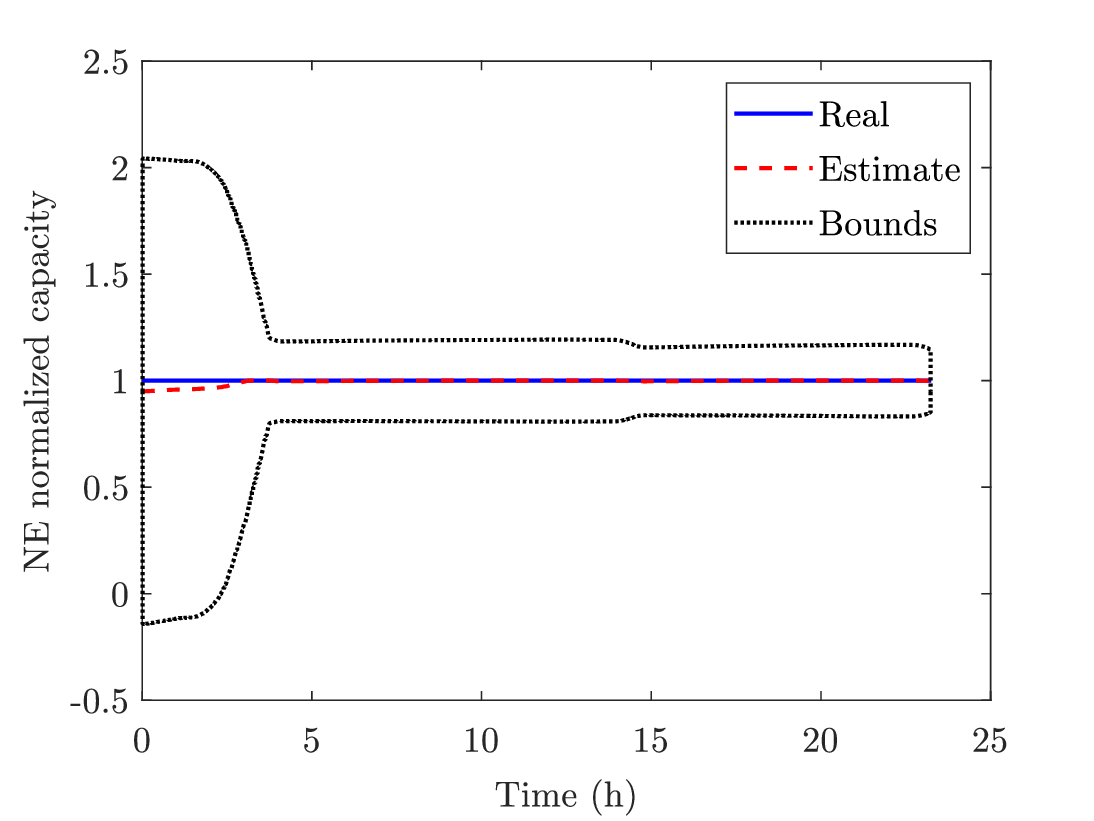}
        \caption{}\label{fig:Val_Exp-SOH_n}
    \end{subfigure}
    
    \begin{subfigure}{0.32\linewidth}
        \centering
        \includegraphics[width=1\textwidth]{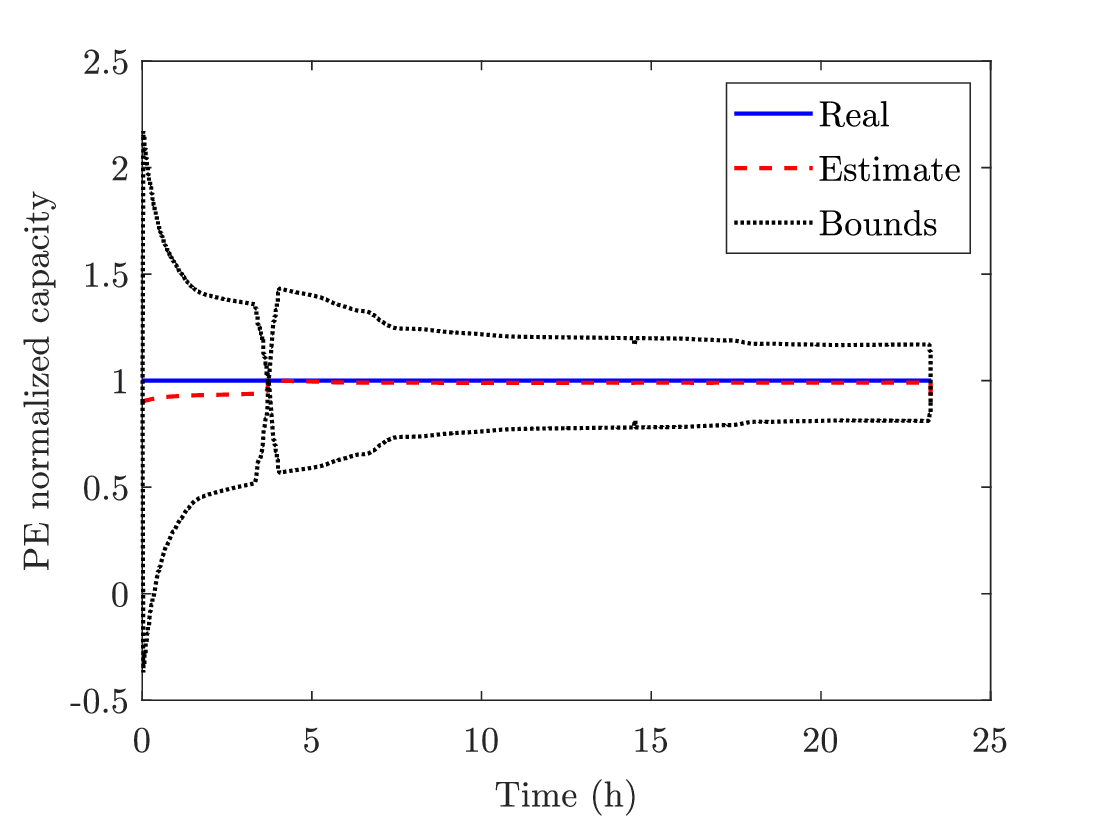}
        \caption{}\label{fig:Val_Exp-SOH_p}
    \end{subfigure}
    \begin{subfigure}{0.32\linewidth}
        \centering
        \includegraphics[width=1\textwidth]{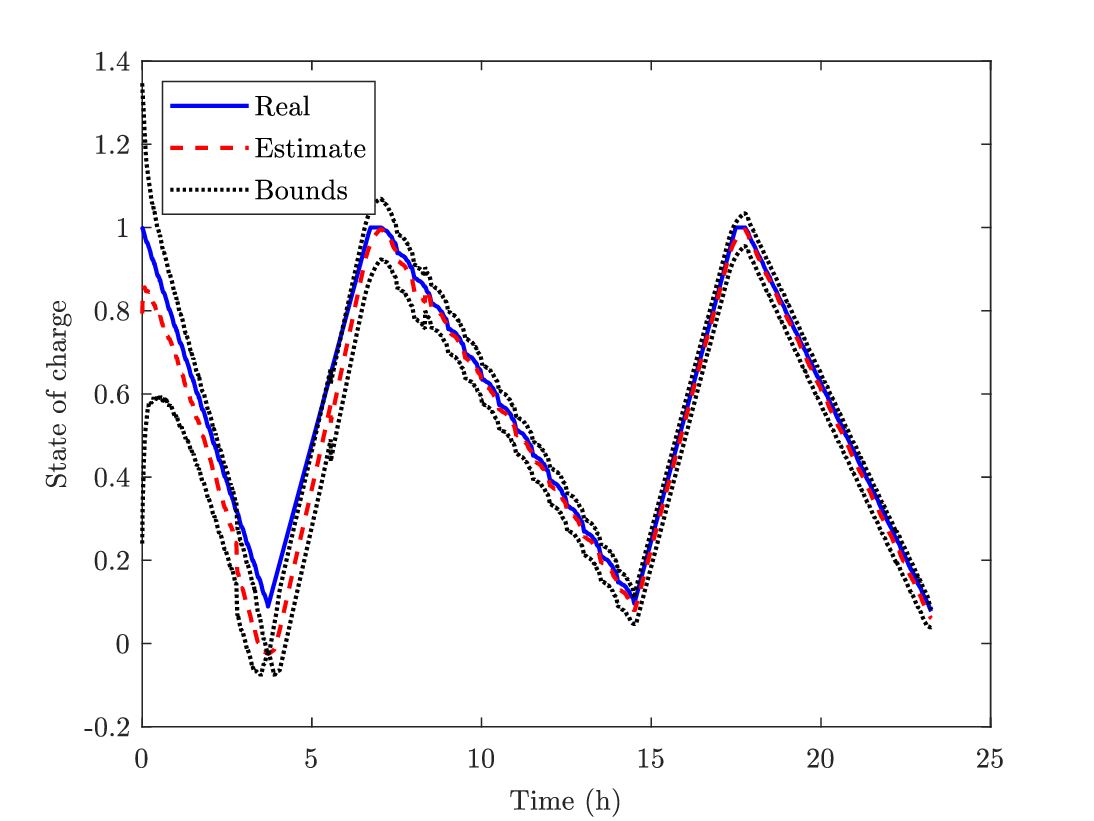}
        \caption{}\label{fig:Val_Exp-SOC}
    \end{subfigure}
    \begin{subfigure}{0.32\linewidth}
        \centering
        \includegraphics[width=1\textwidth]{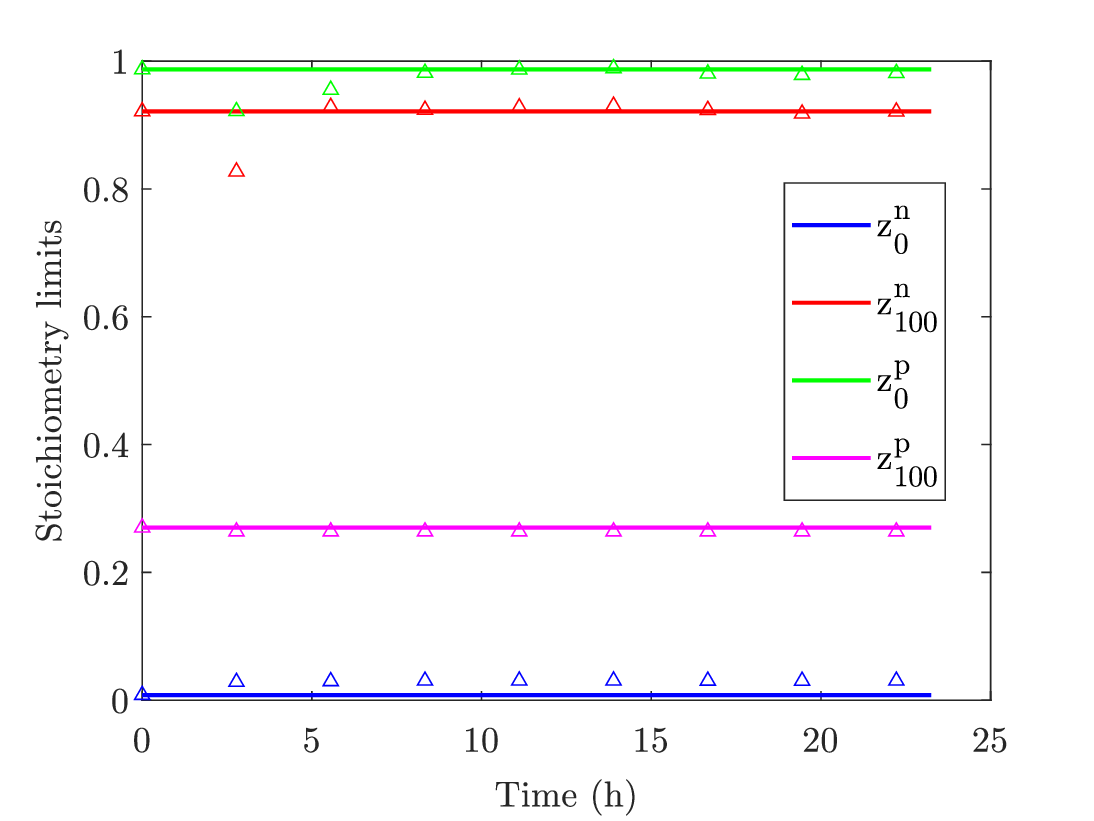}
        \caption{}\label{fig:Val_Exp-Stoichiometries}
    \end{subfigure}
    
    \caption{Estimation of states and parameters for the validation process with a fresh cell. (a)-(b) Negative and positive electrode SOLs, respectively; (c) and (d) normalized negative and positive electrode capacities, respectively; (e) SOC estimate; (f) 100\% and 0\% SOC stoichiometries of both electrodes.}
    \label{fig:Val_Exp}
\end{figure*}

The SOC was initialized with 20\% error, the negative electrode and positive electrode capacities with 5\% and 10\% errors, respectively. The estimation results can be seen in Figure \ref{fig:Val_Exp}. In this case, the ``true'' values were obtained from the previous model characterization, so they might not be perfect values, but the true value should be similar to them. The negative electrode and positive electrode SOLs were obtained using their relation with the SOC and the stoichiometry limits. The SOC itself was calculated using the Coulomb counting method with the laboratory current data.

As shown, the variables are incorrectly initialized, and almost until the end of the first discharge, around 10\% of SOC, they are not totally corrected. In the case of the positive electrode SOL, shown in Figure \ref{fig:Val_Exp-SOL_p}, the variable converges more rapidly than the rest of the variables, since cell voltage is more sensitive to the PE SOL than to the NE SOL or the capacities of both electrodes. By the end of the discharge, around 20\% SOC, the negative electrode SOL, shown in Figure \ref{fig:Val_Exp-SOL_n} converges significantly, since at the end of the discharge the negative electrode OCP has better observability \cite{lopetegi2024}.

As both SOLs converge, the negative and positive electrode capacities approach the correct values as shown in Figures \ref{fig:Val_Exp-SOH_n} and \ref{fig:Val_Exp-SOH_p}. After that, the SOL estimates are accurate for the remainder of the test. Additionally, the stoichiometric window is well calculated with these capacity and SOL values for both electrodes. Finally, the SOC is well estimated once the SOLs, capacities and the stoichiometric windows are accurate, giving accurate electrode-level SOC and SOH estimation values.

\subsubsection{Validation with aged cell}
\begin{figure}[htb!]
        \centering
        \includegraphics[width=0.4\linewidth]{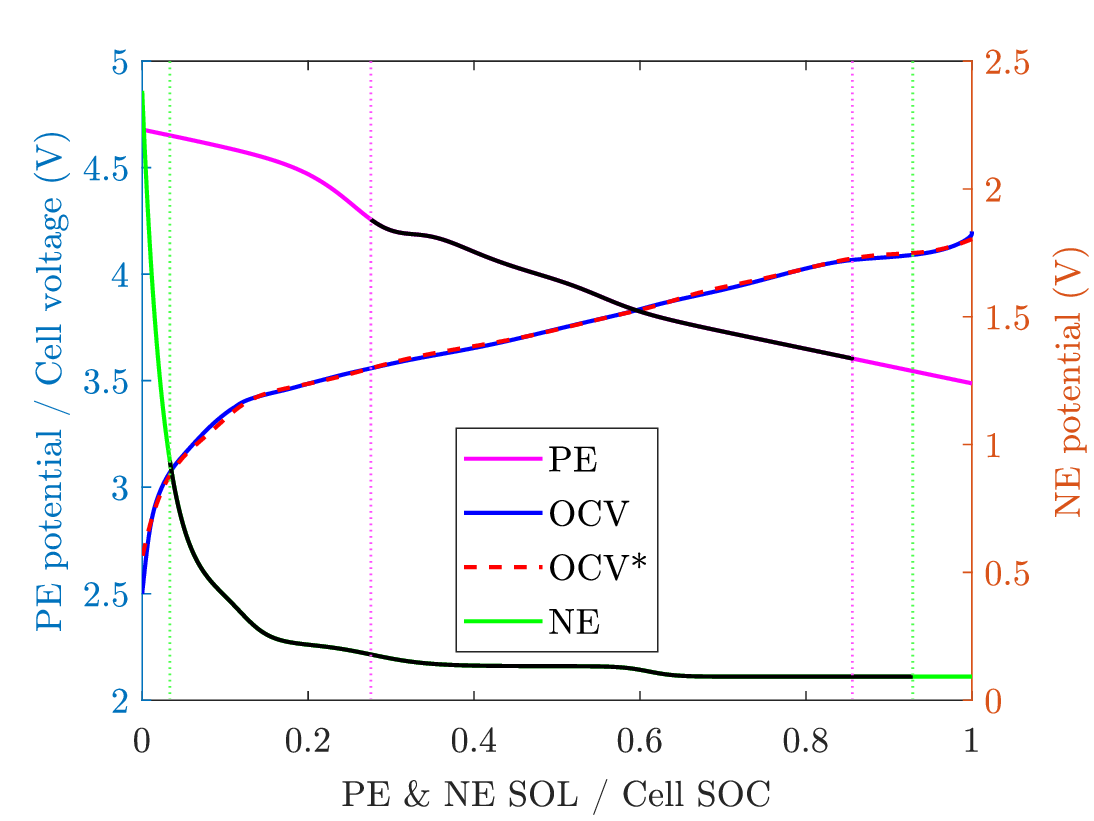}
        \caption{Stoichiometric window fitting from the aged cell OCV data.}\label{fig:ModEst}
\end{figure}

To finish the validation process, consecutive driving cycle experiments were performed on an aged cell with approximately 84\% SOH. The cell was aged for a previous study with constant-current 1C charge and discharge experiments. To obtain the ``true'' experimental degradation mode and stoichiometric window data, the tool ``ModEst'' presented in \cite{SFernandez2024ModEst} was used. Figure \ref{fig:ModEst} shows the fitting result for an OCV test of the aged cell. As can be seen, the fitted OCV agrees well with the experimental data, having a 12 mV RMS error.

Once we obtained this information, both the eSOH and SOC estimates can be validated by considering these fitted parameters as true or measured values.

\begin{figure*}[htb!]
    \centering
    \begin{subfigure}{0.32\linewidth}
        \centering
        \includegraphics[width=1\textwidth]{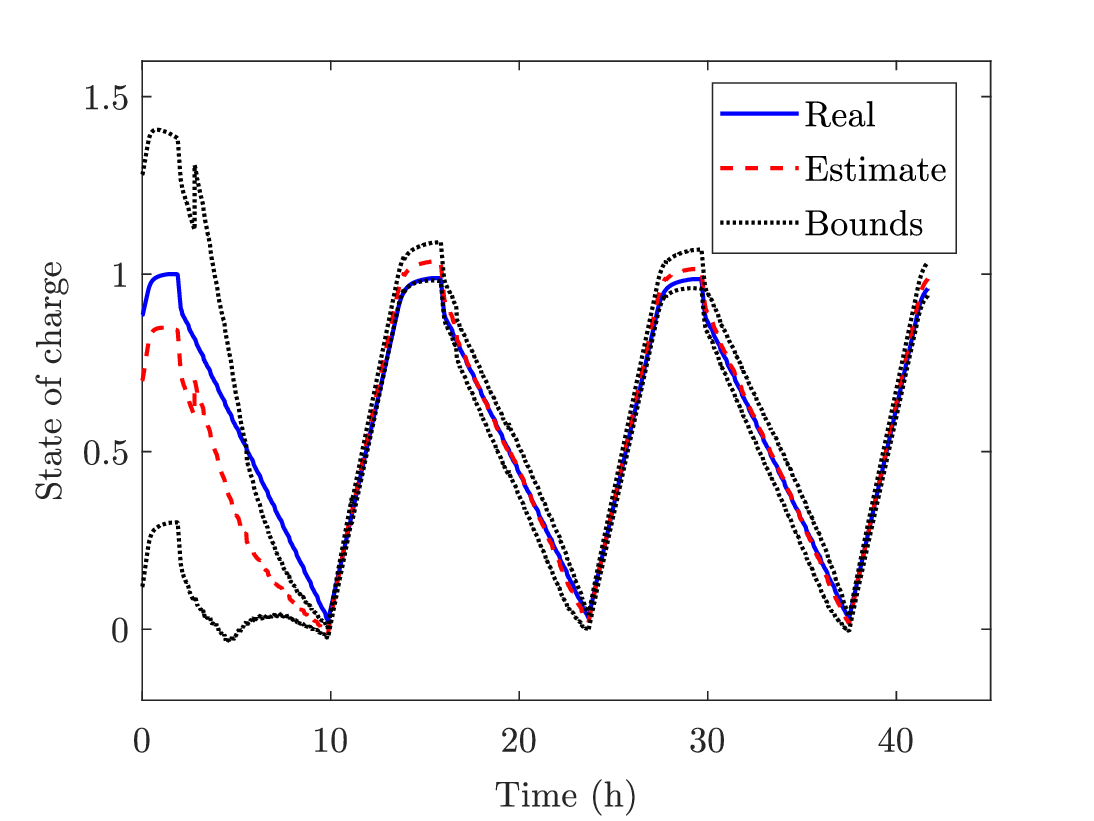}
        \caption{}\label{fig:Val_Exp_Aged-SOC}
    \end{subfigure}
    \begin{subfigure}{0.32\linewidth}
        \centering
        \includegraphics[width=1\textwidth]{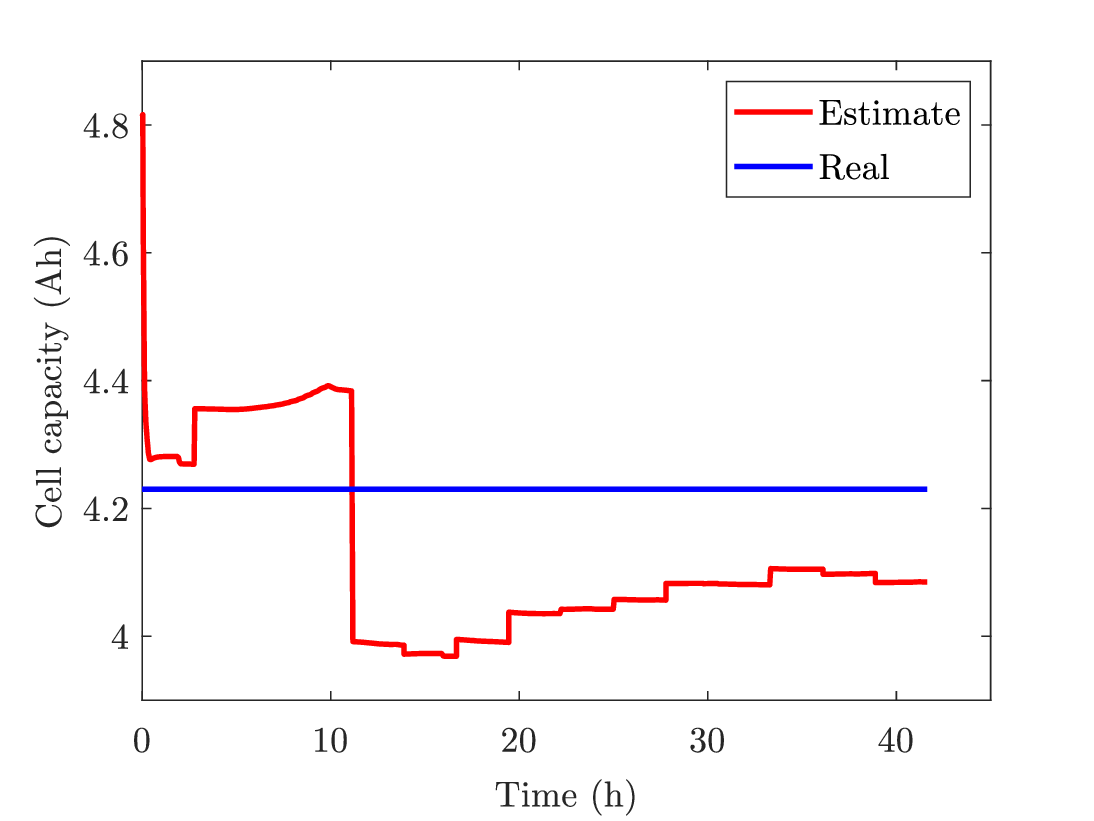}
        \caption{}\label{fig:Val_Exp_Aged-Q}
    \end{subfigure}
    \begin{subfigure}{0.32\linewidth}
        \centering
        \includegraphics[width=1\textwidth]{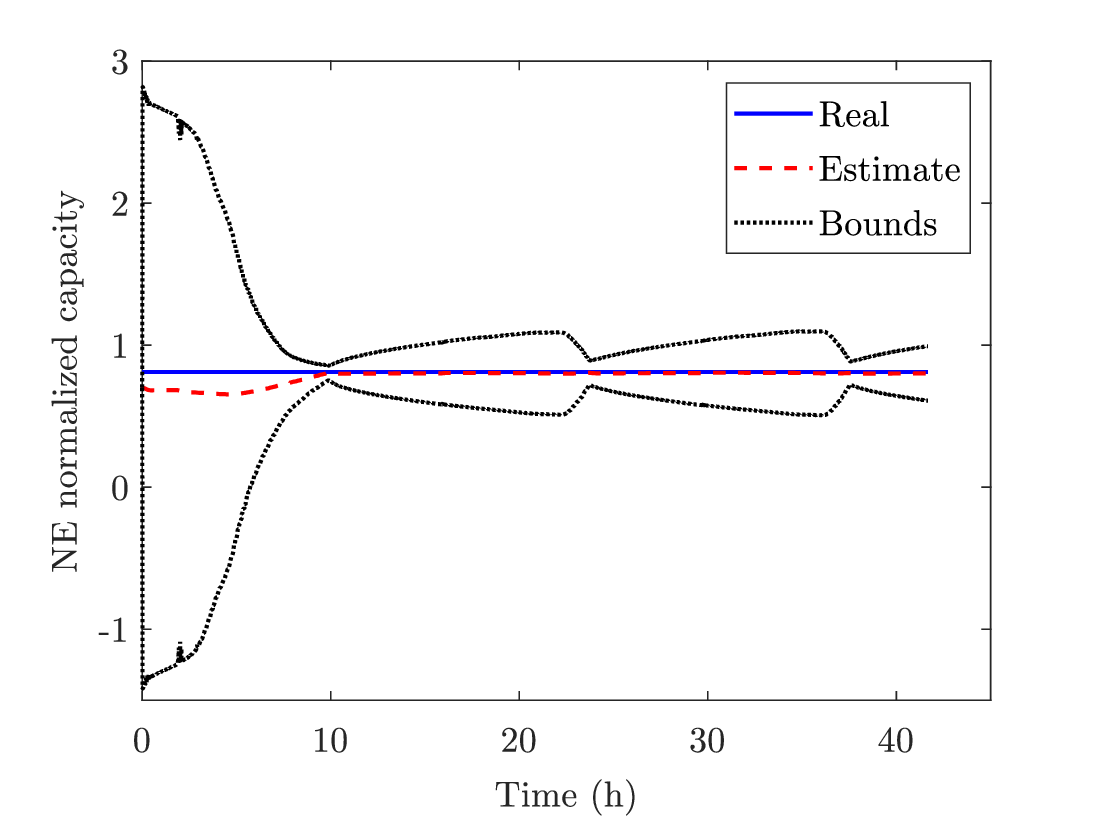}
        \caption{}\label{fig:Val_Exp_Aged-SOH_n}
    \end{subfigure}
    
    \begin{subfigure}{0.32\linewidth}
        \centering
        \includegraphics[width=1\textwidth]{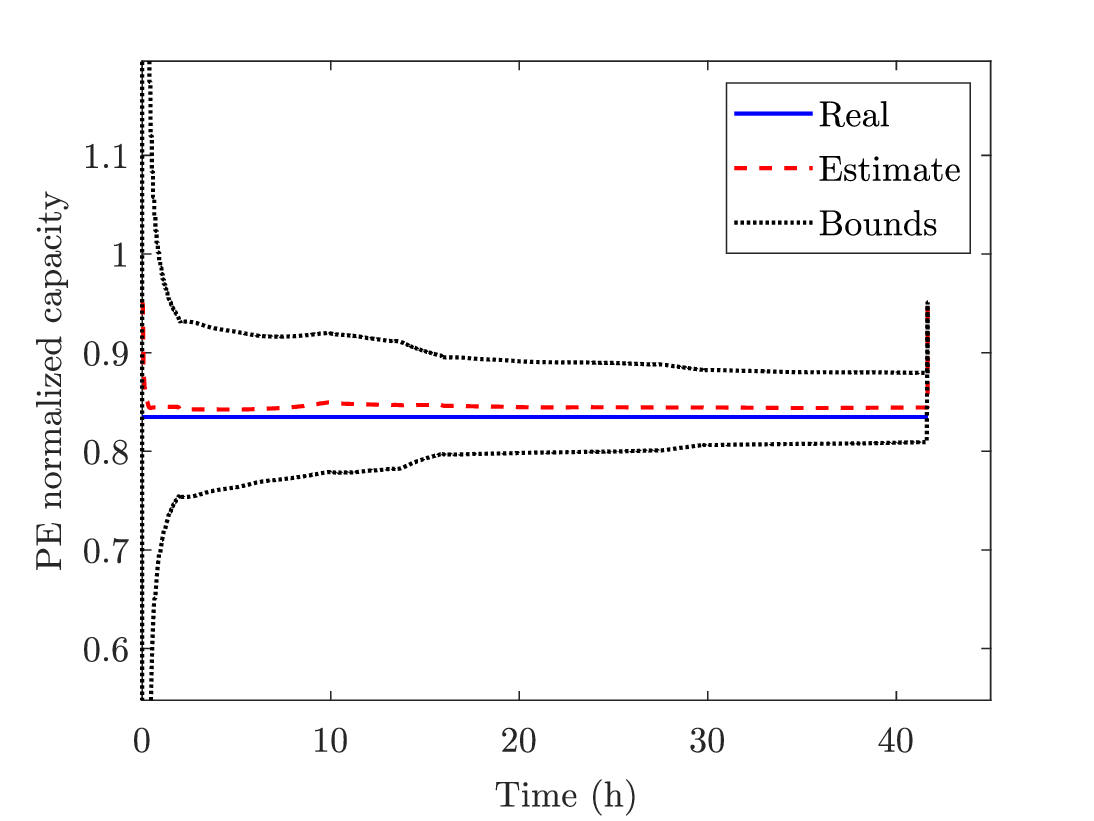}
        \caption{}\label{fig:Val_Exp_Aged-SOH_p}
    \end{subfigure}
    \begin{subfigure}{0.32\linewidth}
        \centering
        \includegraphics[width=1\textwidth]{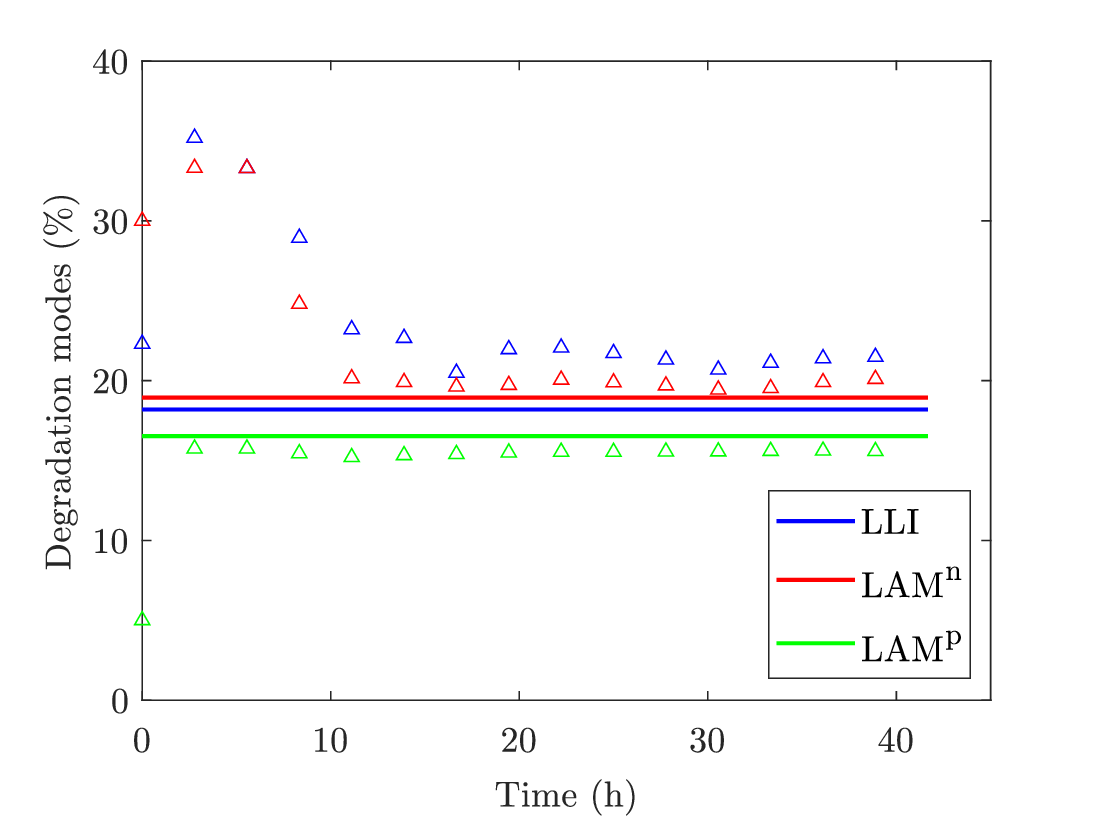}
        \caption{}\label{fig:Val_Exp_Aged-Modes}
    \end{subfigure}
    \begin{subfigure}{0.32\linewidth}
        \centering
        \includegraphics[width=1\textwidth]{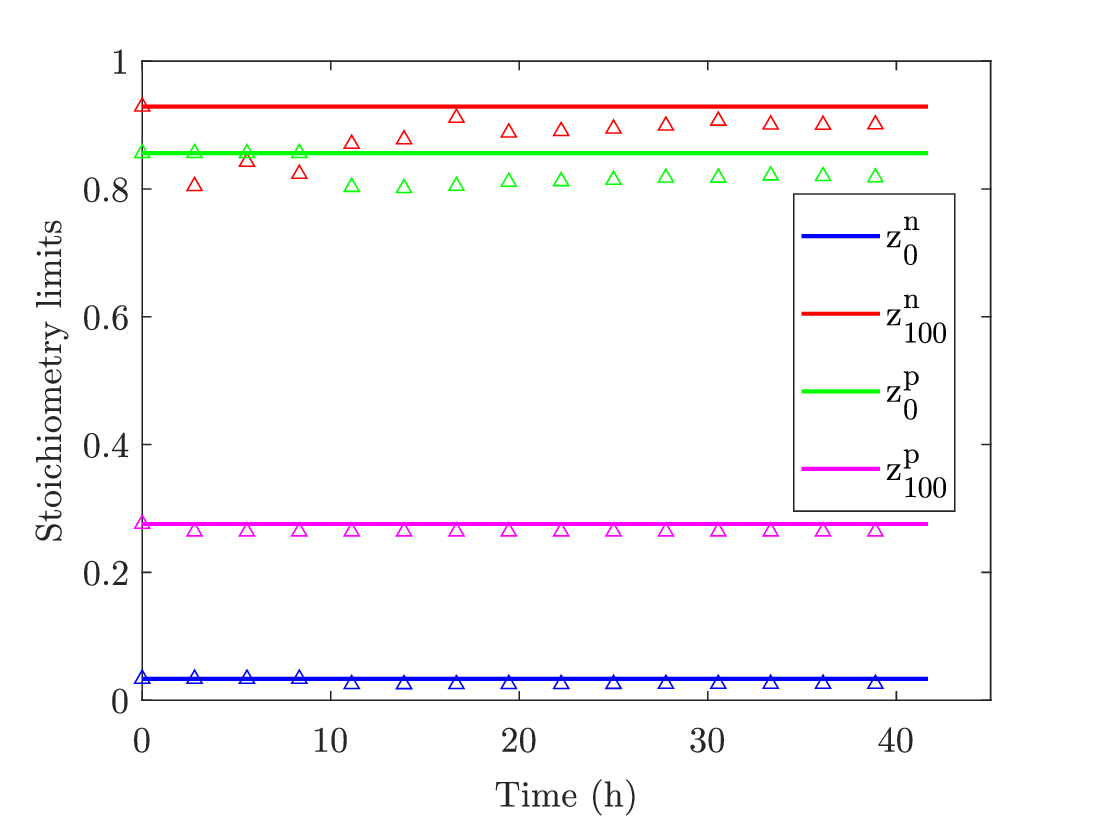}
        \caption{}\label{fig:Val_Exp_Aged-Stoichiometries}
    \end{subfigure}
    
    \caption{Estimation of states and parameters for the validation process with an aged cell. (a) State of charge; (b) Cell capacity; (c) and (d) normalized negative and positive electrode capacities, respectively; (e) Degradation modes; (f) 100\% and 0\% SOC stoichiometries of both electrodes.}
    \label{fig:Val_Exp_Aged}
\end{figure*}

Consecutive UDDS profiles were employed in the tests, as shown in Figure \ref{fig:I_UDDS}. The validation procedure here consists of inserting erroneous electrode capacity and SOC values, as in both above-discussed cases. The results of the state and parameter estimates can be seen in Figure \ref{fig:Val_Exp_Aged}.

 Figure \ref{fig:Val_Exp_Aged-SOH_n} and Figure \ref{fig:Val_Exp_Aged-SOH_p} show the capacity estimates for the negative and positive electrodes, respectively. As can be observed, the AWTLS algorithm is able to correct the capacities, however, the estimates are not as accurate as in the case for BOL in Figure \ref{fig:Val_Exp}. This can be expected due to the additional modeling inaccuracies that might be introduced due to battery aging, as impedance-related parameter changes. Figure \ref{fig:Val_Exp_Aged} shows the SOC estimate, which is well estimated after the capacities converge and the stoichiometric limit values of Figure \ref{fig:Val_Exp_Aged-Stoichiometries} are corrected. Furthermore, with these data we are also able to estimate the cell capacity that gives the traditional capacity-based SOH value. The capacity estimate is shown in Figure \ref{fig:Val_Exp_Aged-Q}. By the end of the test, it has a relative error of approximately 3\%.

 To sum-up, the validation process showed that the algorithm is able to estimate electrode-level SOC and SOH with good accuracy, both in simulation and in experiments.


\section{Conclusions} \label{Conclusions}

This work presented a novel method to estimate electrode-level SOC and SOH from typical operation data using dynamic cycling profiles, based on interconnected SPKFs and an eECM. With the obtained data, all specified degradation modes and the SOH value can be obtained. It was shown that the estimator works accurately using both simulation and experimental tests. However, it is worth noting that the estimator converges more reliably to the true values when the state variables have better observability, as when the negative electrode OCP exhibits highly nonlinear behavior. Therefore, depending on the test SOC range, better or worse estimates could be obtained. Future work will investigate the estimation accuracy with different input current profiles and SOC ranges.

Regarding the eECM, one of the important aspects to detect battery degradation more accurately, was to obtain resistance and capacitance values for a wide SOL range, well beyond the initial SOC range. For that, the eECM was characterized using a PBM. This way, the resistor and capacitor values were obtained for wider SOL windows. This should improve accuracy compared to the SOC method when there is degradation in the battery, and SOLs that were not achieved at BOL are reached, or the SOLs shift with the SOCs. In this regard, another goal of our future work will be to obtain a reliable, fast, and cheap method to obtain an eECM, since it should not be necessary to build a PBM in order to obtain an eECM. 

\section{Acknowledgement}

This research did not receive any specific grant from funding agencies in the public, commercial, or not-for-profit sectors.


\bibliographystyle{unsrt}
\bibliography{references}


\appendix

\section{PBM parameters}\label{ap:PBMparams}

The model parameters used for the simulations correspond to an LG M50 cell with NMC/graphite-SiO\textsubscript{x} chemistry and were acquired from \cite{chen2020}. The parameters are shown in Table \ref{tab:params}. The functions for the negative and positive open circuit potentials (OCPs) are given by
\begin{eqnarray}
U_{ocp}^n(\theta) = 1.9793e^{-39.3631\theta} + 0.2482 \cr
- 0.0909 \times \tanh(29.8538(\theta-0.1234)) \cr
- 0.04478 \times \tanh(14.9159(\theta-0.2769)) \cr
- 0.0205 \times \tanh(30.4444(\theta-0.6103))
\end{eqnarray}
and
\begin{eqnarray}
U_{ocp}^p(\theta) = -0.809\theta + 4.4875 \cr
- 0.0428 \times \tanh(18.5138(\theta-0.5542)) \cr
- 17.7326 \times \tanh(15.789(\theta-0.3117)) \cr
+ 17.5842 \times \tanh(15.9308(\theta-0.312))
\end{eqnarray}
respectively. The ionic conductivity of the electrolyte is given by
\begin{eqnarray}
    \kappa_e(c_e) = 1.297 \times 10^{-10}c_e^3 - 7.94 \times 10^{-5}c_e^{1.5} \\ + 3.329 \times 10^{-3}c_e,
\end{eqnarray}
and the electrolyte diffusivity $D_e(c_e)$ by
\begin{eqnarray}
    D_e(c_e) = 8.794 \times 10^{-17}c_e^2 - 3.972 \times 10^{-13}c_e \\ + 4.862 \times 10^{-10}.
\end{eqnarray}

\begin{table*}[htb!]
\centering
\caption{\label{tab:PBMparams} LG M50 cell parameters \cite{chen2020}.} 
\resizebox{\textwidth}{!}{%
    \begin{tabular}{@{}*{5}{l}}
    \toprule                              
    Parameter & Description &Negative&Separator&Positive\cr 
    \midrule
    $D_s$ (m\textsuperscript{2} s\textsuperscript{-1}) & Solid-phase diffusivity & $6.069\times10^{-13}$ & & $1.225 \times 10^{-14}$  \cr
    $\sigma$ (S m\textsuperscript{-1}) & Electric conductivity &215& & 0.18 \cr 
    $c_{s,max}$  (kmol m\textsuperscript{-3})& Maximum solid-phase concentration &33.133 & & 63.104 \cr 
    $c_{e,0}$ ( kmol m\textsuperscript{-3}) & Initial electrolyte concentration & 1 &1 &1 \cr
    $A$ (m\textsuperscript{2}) & Area & $0.1027$ & $0.1027$ & $0.1027$ \cr
    $L$ (m) & Thickness &$8.52 \times 10^{-5}$ & $1.2 \times 10^{-5}$ & $7.56 \times 10^{-5}$ \cr 
    $R_s$ (m) & Particle radius & $5.86 \times 10^{-6}$ & & $5.22 \times 10^{-6}$ \cr 
    $\alpha$ & Charge transfer coefficient & 0.5 & & 0.5 \cr
    $R_{f} (\Omega)$ & Film resistance & 0.02 & & 0 \cr
    $\varepsilon_s$ & Active material volume fraction& 0.75 & & 0.665 \cr
    $\varepsilon_e$ & Porosity & 0.25 & 0.47 & 0.335 \cr
    $brug$ & Bruggeman's exponent & 1.5 & 1.5 & 1.5 \cr
    $z_{0 \%}$ & Stoichiometry at 0 \% SOC & 0.008 & & 0.987 \cr
    $z_{100 \%}$ & Stoichiometry at 100 \% SOC & 0.9214 & & 0.27 \cr
    $k_{0,norm}$ (mol m\textsuperscript{-2} s\textsuperscript{-1}) & Normalized reaction rate& $7.04 \times 10^{-6}$ & & $7.07 \times 10^{-5}$ \cr
    $t_+^0$ & Transference number & 0.2594 & 0.2594 & 0.2594 \cr
    \bottomrule
    \end{tabular}}
\end{table*}

\section{NE passive circuit elements parametrization} \label{ap:NE_parametriztion}

\begin{figure}[htb!]
    \centering
    \includegraphics[width=0.4\linewidth]{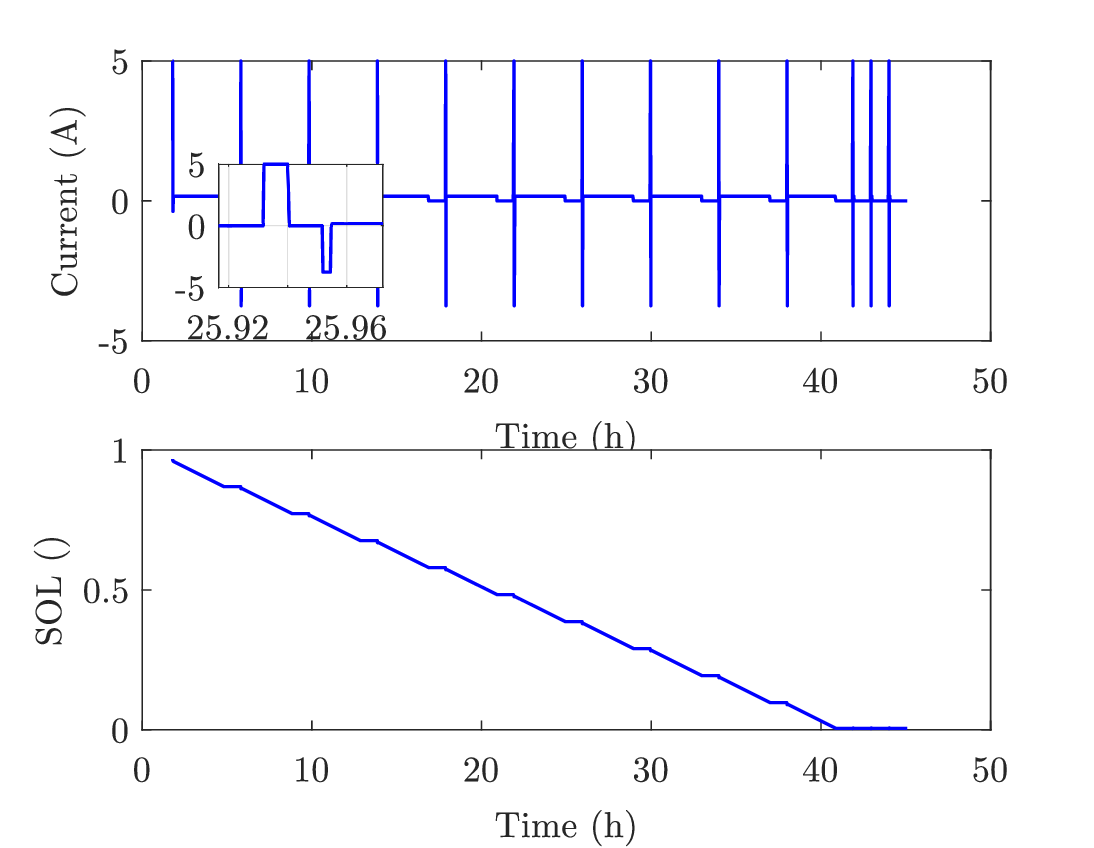}
    \caption{(a) HPPC current profile and (b) NE SOL evolution during test.}
    \label{fig:ParamFit:HPPCneg}
\end{figure}

For the NE passive circuit element values parametrization, an HPPC test is performed in a sufficiently wide SOL range, covering a wider SOL range than stoichiometry limits, as shown in Figure \ref{fig:ParamFit:HPPCneg}.

\begin{figure}[htb!]
    \centering
    \subfloat[\centering\label{fig:ParamFit:Vneg}]{{\includegraphics[width=0.4\linewidth]{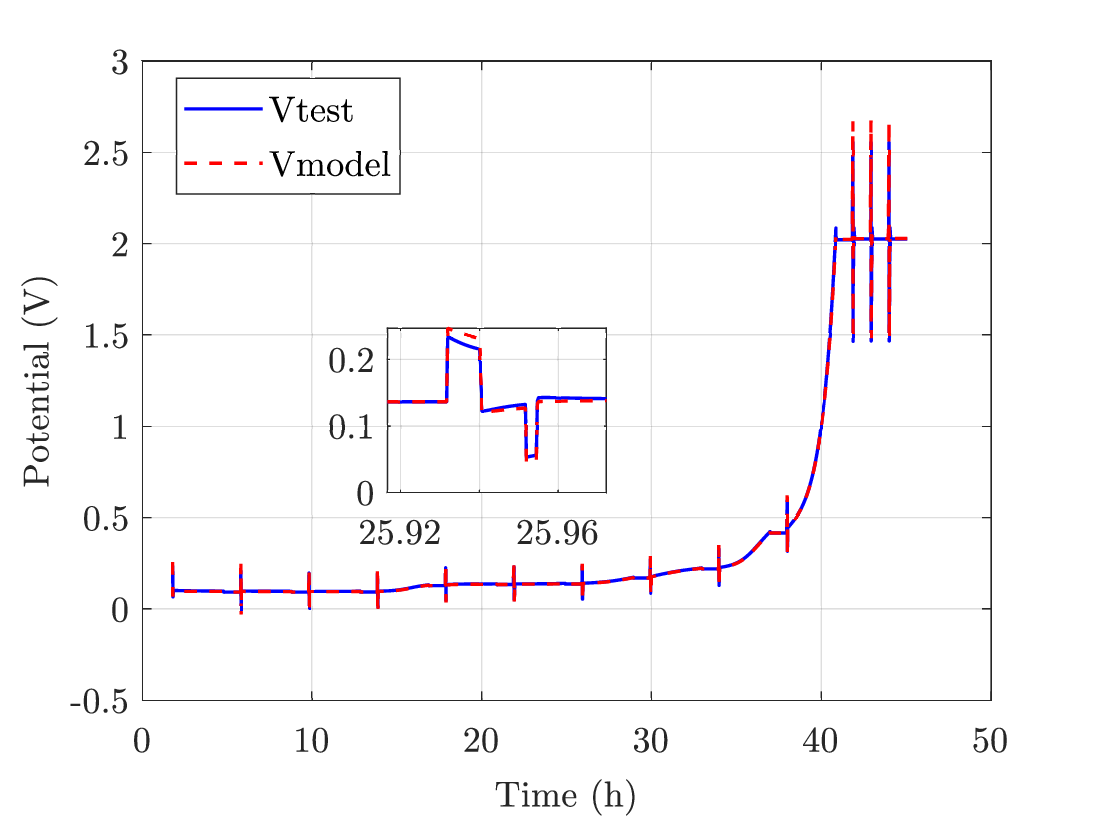}}}\\
    \subfloat[\centering\label{fig:ParamFit:DVneg}]{{\includegraphics[width=0.4\linewidth]{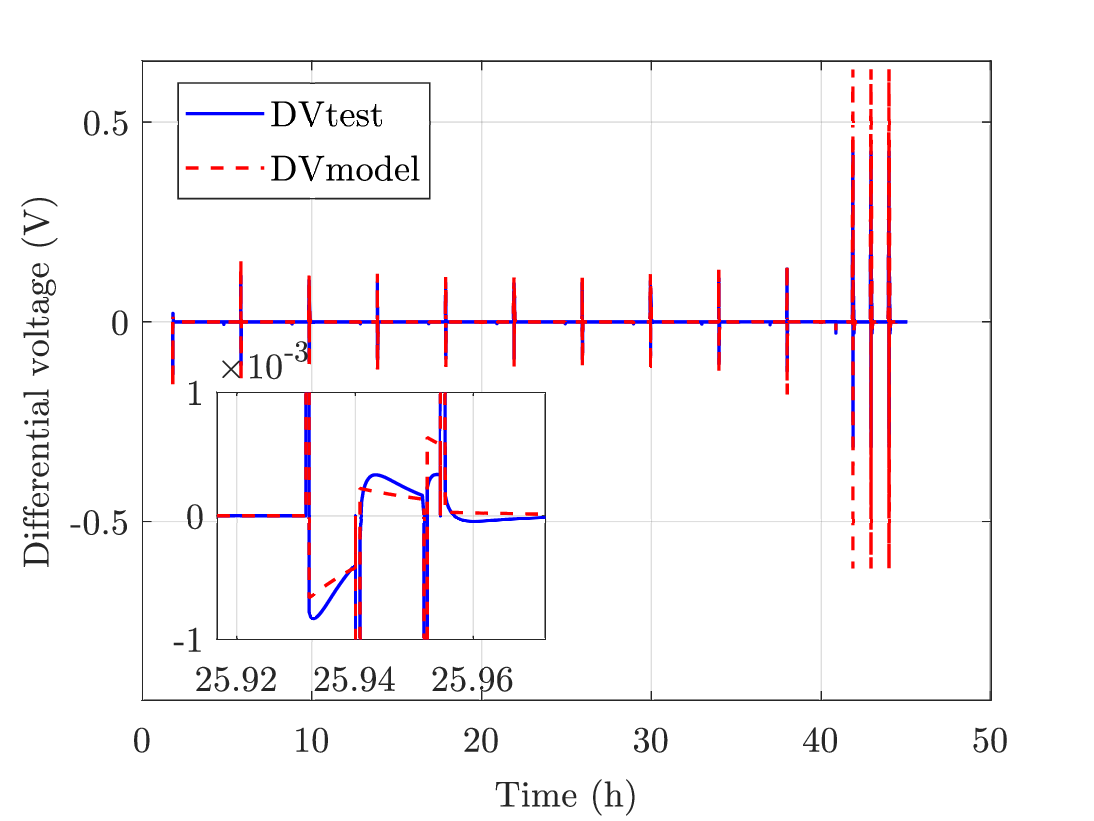}}}
    \caption{NE model (a) output voltage and test voltage comparison; (b) differential voltage response comparison.}
    \label{fig:ParamFitOutput_neg}
\end{figure}

The obtained element values are shown in Table \ref{tab:HCparams} and the voltage and differential voltage response compared to the curves obtained at test is presented in Figure \ref{fig:ParamFitOutput_neg}.

\end{document}